\documentclass{aa}  
\usepackage{natbib}
\usepackage{graphicx}
\usepackage[x11names]{xcolor}
\usepackage{supertabular}

\usepackage[varg]{txfonts}
\usepackage{siunitx}
\DeclareSIUnit\um{\micro\meter}
\DeclareSIUnit\Msun{M_{$\odot$}}

\newcommand{\cdbox}[1]{%
  \colorlet{currentcolor}{.}%
  {\color{DodgerBlue2}%
    \dbox{\color{currentcolor}#1}}%
}
\usepackage{dashbox,framed,color,ocg-p}
\newcommand{\ToggleLayer}[2]{%
  \leavevmode
  \pdfstartlink user {
    /Subtype /Link
    /Border [0 0 0]%
    /A <<
      /S/JavaScript
      /JS (
         var aOCGs = this.getOCGs(), Layer;
         var Layers = "#1".split(","), Active = -1, i, l;
         for (l=0; l<Layers.length; l++) {
           Layer = Layers[l];
           for (i=0; aOCGs && i<aOCGs.length; i++) {
             if (aOCGs[i].state && aOCGs[i].name == Layer) {
               Active = l;
               aOCGs[i].state = false;
             }
           }
           if (Active >= 0) break;
         }
         if (Active == -1) {
           for (l=0; l<Layers.length; l++) {
             if (Layers[l] ==   ) Active = l;
           }
         }
         Active = Active + 1;
         if (Active == Layers.length) Active = 0;
         Layer = Layers[Active];
         for (i=0; aOCGs && i<aOCGs.length; i++) {
           if (aOCGs[i].name == Layer) aOCGs[i].state = true;
         }
      )
    >>
  }#2%
  \pdfendlink
}

\title{\emph{Herschel-Planck} dust optical depth and column density maps}
\subtitle{II. Perseus.} 
\author{Eleonora Zari \inst{1,2}
\and Marco Lombardi \inst{2}
\and  Jo\~ao Alves \inst{3}
\and Charles J. Lada \inst{4}
\and Hervé Bouy \inst{5}
}
\institute{
{1} Leiden Observatory, Niels Bohrweg 2, 2333 CA Leiden, the Netherlands;
{2} University of Milan, Department of Physics, via Celoria 16, 20133 Milan, Italy;
{3} University of Vienna,  Turkenschanzstrasse 17, 1180 Vienna, Austria;
{4} Harvard-Smithsonian Center for Astrophysics, Mail Stop 72, 60 Garden Street, Cambridge, MA 02138;
{5} Centro de Astrobiolog\'ia, INTA-CSIC, PO Box 78, 28691 Villanueva de la Canada, Madrid, Spain}

\begin{document}

\abstract{We present optical depth and temperature maps of the Perseus molecular cloud, obtained combining dust emission data from the \emph{Herschel} and \emph{Planck} satellites and 2MASS/NIR dust extinction maps. The maps have a resolution of 36~arcsec in the \emph{Herschel} regions, and of 5~arcmin elsewhere. The dynamic range of the optical depth map ranges from $1\times10^{-2}\, \mathrm{mag}$ up to $20 \,\mathrm{mag}$ in the equivalent K band extinction. We also evaluate the ratio between the \SI{2.2}{\um} extinction coefficient and the \SI{850}{\um} opacity. The  value we obtain is close to the one found in the Orion B molecular cloud. We show that the cumulative and the differential area function of the data (which is proportional to the probability distribution function of the cloud column density) follow power laws with index respectively $\simeq -2$, and $\simeq -3$. We use WISE data  to improve current YSO catalogs based mostly on \emph{Spitzer} data and we build an up-to-date selection of Class~I/0 objects. Using this selection, we evaluate the local Schmidt law, $\Sigma_{\mathrm{YSO}} \propto \Sigma_{\mathrm{gas}}^{\beta}$, showing  that $\beta=2.4 \pm 0.6$.   
Finally, we show that  the area-extinction relation is important for determining the star formation rate in the cloud, which is in agreement with other recent works.}

\keywords{ISM: clouds, dust, extinction, ISM: structure, ISM: individual objects: Perseus molecular cloud, Methods: data analysis} 

\maketitle

\section{Introduction}
Understanding the physical processes that rule star formation is fundamental to improve our knowledge of galaxy formation and evolution across cosmic time. 

An essential step towards reaching a better comprehension of star formation inside our Galaxy is to establish the relationship between the star formation rate (SFR) and the physical properties of the interstellar gas. 
\citet{Schmidt1959} conjectured that "the SFR depends on the gas density and [...] varies with a power law of the gas density." Schmidt argued that the index of the power law in the solar neighborhood was $\sim 2$. \citet{Kennicutt1998} expressed the law as $\Sigma_{\mathrm{SFR}} = \kappa \Sigma_{\mathrm{gas}}^{\beta}$, where $\Sigma_{\mathrm{gas}}$ and $\Sigma_{\mathrm{SFR}}$ are the surface densities of the gas and of star formation rate.  Kennicutt tested the Schmidt law in a heterogeneous sample of star forming galaxies and showed that a power law scaling relation exists between them and is characterized by an index $\beta = 1.6$. 
Using observations of local disc and starburst galaxies, \citet{Bigiel2008} found a complex scaling between $\Sigma_{\mathrm{SFR}}$ and $\Sigma_{\mathrm{gas}}$, which could not be described by a single power law. On smaller, sub-kpc scales, \citet{Lada2013} showed that there is no Schmidt scaling relation between local giant molecular clouds, but a series of observational studies demonstrated that a Schmidt scaling law exists within some local molecular clouds \citep{Heiderman2010, Gutermuth2011, Lombardi2013, Lada2013, Evans2014, Harvey2014}.

In order to be able to understand the detailed interplay between the structure of molecular clouds and their \emph{local} star formation rates (as opposed to the global one often considered), we are carrying out a coordinated study of molecular clouds in the Gould belt.  We use a combination of \emph{Planck} and \emph{Herschel} dust emission data, calibrated using near infrared (NIR) dust extinction.  This technique guarantees an optimal resolution of the cloud column density maps (corresponding to the \emph{Herschel}/SPIRE500 \SI{36}{arcsec} resolution), better than the one of standard NIR extinction maps in the region of the sky we are considering.  
In our first paper we considered the Orion giant molecular cloud \citep{Lombardi2014}; here we focus on Perseus.

The Perseus cloud was first observed by Barnard, who cataloged a portion of it as Barnard 1-5 and Barnard 202-206. Since then, the cloud was extensively studied using molecular line emission \citep{Ridge2006}, star count extinction \citep{Bachiller1984}, and dust continuum emission \citep{Hatchell2005, Kirk2006, Enoch2006}. 
Distance estimates for Perseus vary \citep{{Herbig1983}, {Cernis1990}, {Cernis1993}, {Bally2008}, {Hirota2008}, {Schlafly2014}}; we use \SI{240}{pc} for consistency with \citet{Lombardi2010}.
Perseus is the prototype intermediate mass star forming region, with young B stars and two clusters, IC 348  \citep{Muench2007} and NGC1333 \citep{Lada1996}. The complex seems to be associated with the Perseus OB2 association \citep{deZeeuw1999}. 
Recently, \emph{Herschel} observations were used to characterize small regions of the cloud. \citet{Pezzuto2012} studied two star forming dust condensations, B1-bS and B1-bN in the B1 region (see Fig.~\ref{fig:fig02}). They conclude that these two sources could be good examples of the first hydrostatic core phase. 
\citet{Sadavoy2012} presented observations of the B1-E region and proposed that it may be forming the first generation of dense cores.  
Finally, \citet{Sadavoy2014} identified and characterized 28 candidate Class~0 protostars in the whole cloud, four of which are newly discovered. They also found that the star formation efficiency of the clumps, as traced by Class~0 protostars, correlates strongly with the flatness of their respective column density distribution at high values.    
A global study of the Perseus cloud properties with \emph{Herschel} data is still missing, and constitutes the main goal of this paper.

This paper is organized as follows. In Section \ref{Data} we present the data used in this work; in Section \ref{Method} we briefly describe the data reduction process; in Section \ref{Results} we apply the method to study the Perseus molecular cloud and present the column density and temperature maps \footnote{The maps will be publicly available through CDS (http://cdsweb.u-strasbg.fr) and the website http://www.interstellarclouds.org}; in Section \ref{Section:Schmidt_law} we derive the local Schmidt law; finally, in Section \ref{Conclusions} we present a summary.

\section{Data\label{Data}}
Perseus was observed as part of the \emph{Gould Belt Survey} \citep{Andre2010}, one of the \emph{Herschel} satellite key projects. The cloud was observed using the two photometers PACS \citep{Poglitsch2010} and SPIRE \citep{Griffin2010}, in five bands, centered approximately at \SIlist{70;160;250;350;500}{\um}. Table \ref{Tab:Herschel_obs} gives an overview of the observation used in this work.
Dust maps derived by the \emph{Planck} Collaboration \citep{Planck2014}, and the near infrared extinction maps produced by \citet{Lombardi2010} through the \textsc{NICEST} method \citep{Lombardi2009} are also used. We selected the region (corresponding to the Perseus molecular cloud) with galactic coordinates:
\begin{equation}
155^{\circ} < l < 165^{\circ}, \hspace{0.5cm}  -25^{\circ}< b < -15^{\circ}.
\end{equation}
Additionally, we employed the AllWISE data release \citep{2010AJ....140.1868W, Mainzer2011} to select Young Stellar Objects (YSOs) from their colors. 
The WISE satellite observed the whole sky in four infrared bands, often referred to as W1, W2, W3, and W4, with wavelengths centered at \SIlist{3.4;4.6;12;22}{\um}. 
We retrieved 1.43 million sources from the WISE point source catalog  in the science field and 1.63 million sources in the control field.  We rejected any source with contamination and confusion flags and we further restricted the selection to those measurements with errors $\sigma < 0.15 \, \mathrm{mag}$ in all the WISE bands.
The  AllWISE source catalog contains associations with the 2MASS Point Source Catalog \citep{2006AJ....131.1163S}. The precision adopted by the AllWISE collaboration for the 2MASS association is 3". In addition, we required photometric errors in the 2MASS bands to be smaller than $0.1\,\mathrm{mag}$ .
\begin{table*}
\caption{\emph{Herschel} parallel mode observation used.}
\begin{centering}
\begin{tabular}{ccccccc}
Target Name & Obs. ID & RA & DEC & Wavelengths ($\mu m$) & Obs. date & Exp. Time (s) \tabularnewline
\hline
\tabularnewline
Perseus S & 1342190326 & 3h 29m 38s & +30 54' 33" & SPIRE, PACS 70 and 160 & 2010-09-02 &  14765 \tabularnewline
Perseus N & 1342214504 & 3h 42m 51s & +32 01' 38" & SPIRE, PACS 70 and 160 & 2011-19-2 & 11624 
\end{tabular}
\par\end{centering}
\label{Tab:Herschel_obs}
\end{table*}

\section{Method\label{Method}}
The reduction technique is very close to the one presented by  \cite{Lombardi2014} and therefore we only briefly summarize it in this section.

\subsection{Dust Model}
Dust is optically thin at the \emph{Herschel} observation frequencies (at least for $\lambda > $ \SI{160}{\um}), and therefore its emission can be modeled as a modified black body:
\begin{equation}\label{eq:emission}
I_{\nu}=B_{\nu}\left(T\right)\left(1-e^{-\tau_{\nu}}\right) \simeq B_{\nu}\left(T\right)\tau_{\nu}.
\end{equation}
Here $B_{\nu}\left(T\right)$ is the black body function at the temperature $T$ and the optical thickness $\tau_{\nu}$ is taken to be a power law of the frequency $\nu$:
\begin{equation}
\tau_{\nu}= \tau_{\nu_0}\left(\frac{\nu}{\nu_0}\right)^{\beta}.
\end{equation}
The frequency $\nu_0$ is an arbitrary reference frequency that we set as $\nu_0 = 353 \, \mathrm{GHz}$ (corresponding to $\lambda = \SI{850}{\um}$), similarly to what was done by the \emph{Planck} collaboration \citep{Planck2014}.

\begin{figure*}
  \centering
  \includegraphics[width=\hsize]{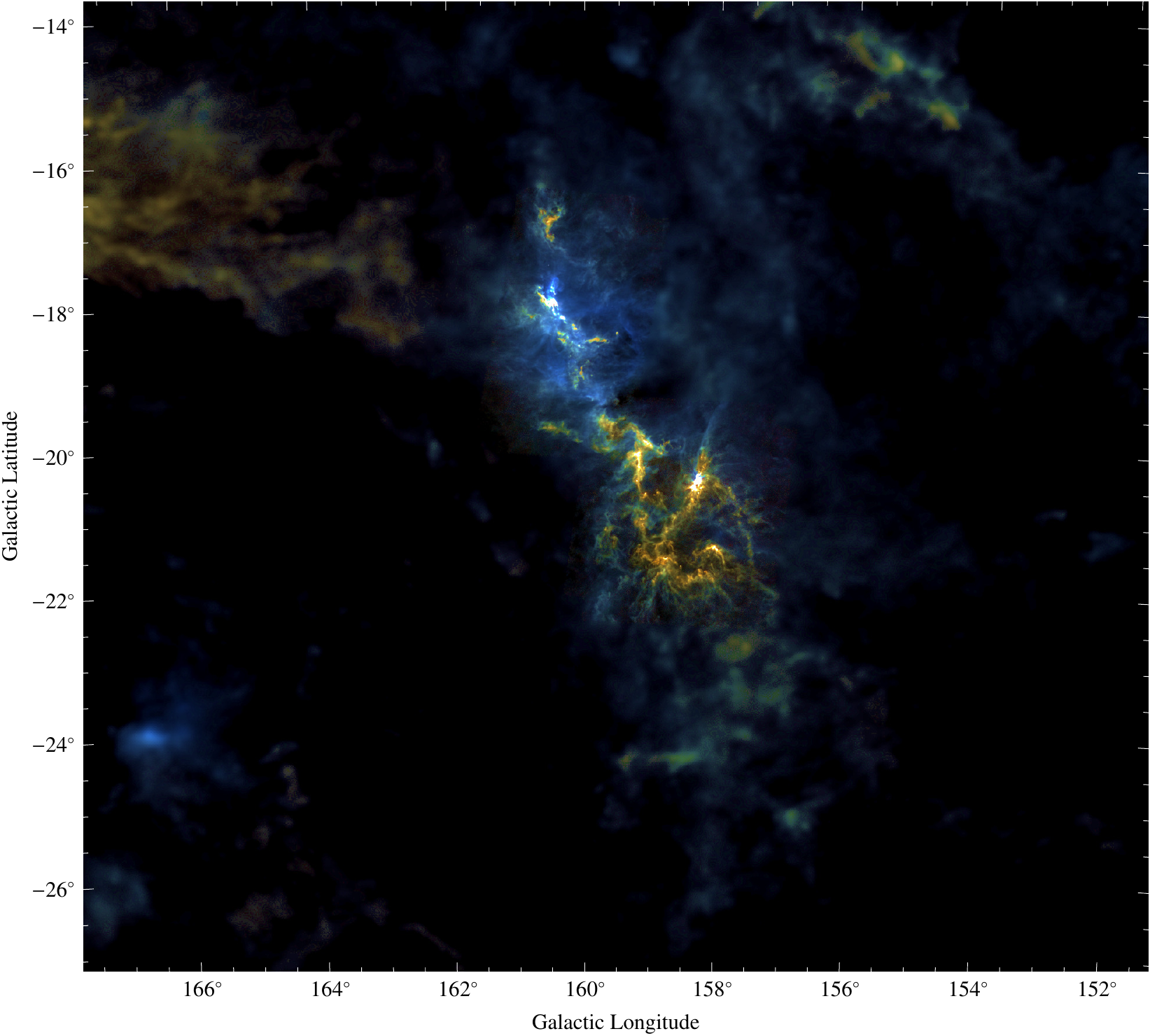}%
  \hspace{-\hsize}%
  \begin{ocg}{fig:2a}{fig:2a}{0}%
  \end{ocg}%
  \begin{ocg}{fig:2b}{fig:2b}{1}%
    \includegraphics[width=\hsize]{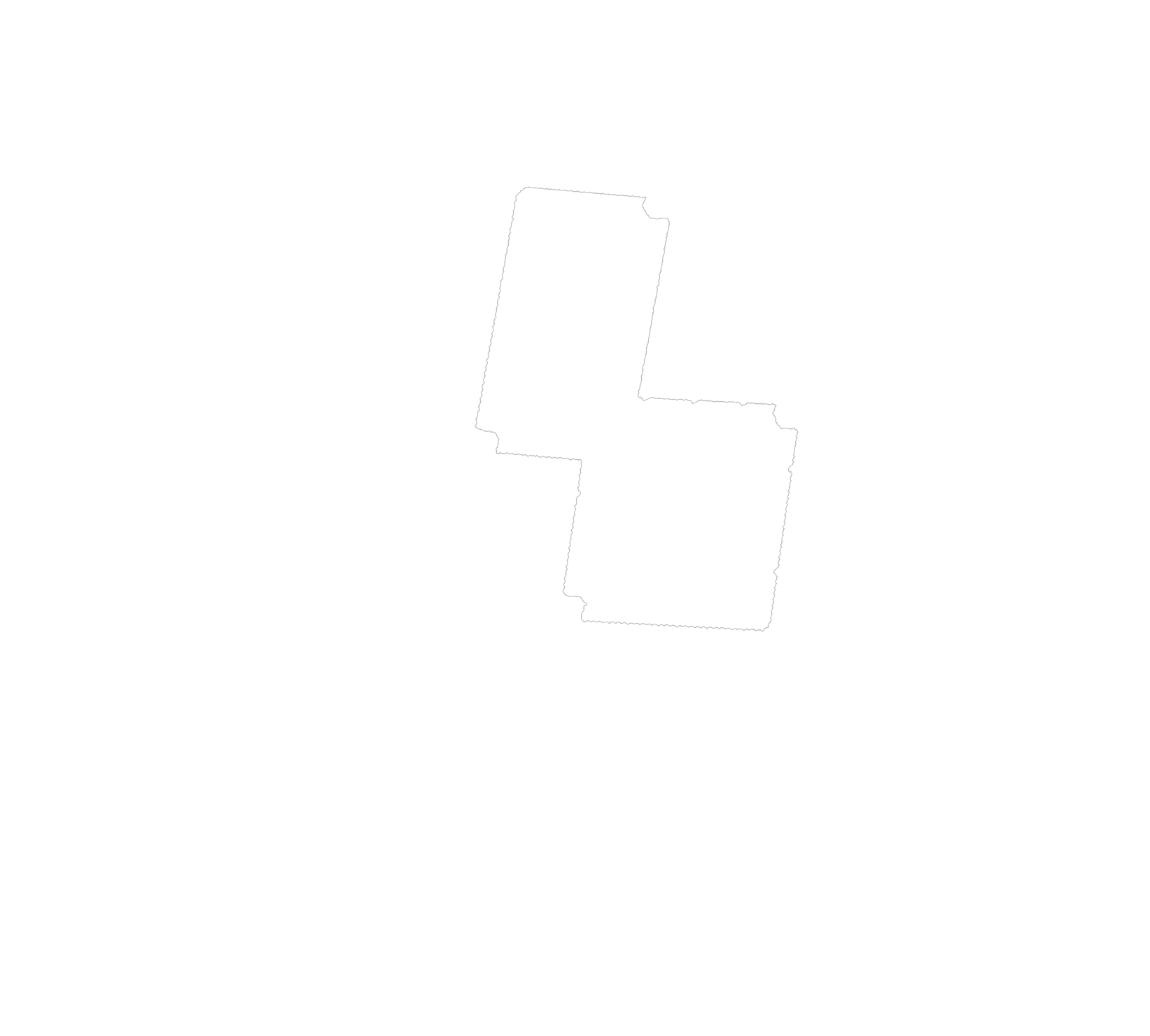}%
    \hspace{-\hsize}%
    \includegraphics[width=\hsize]{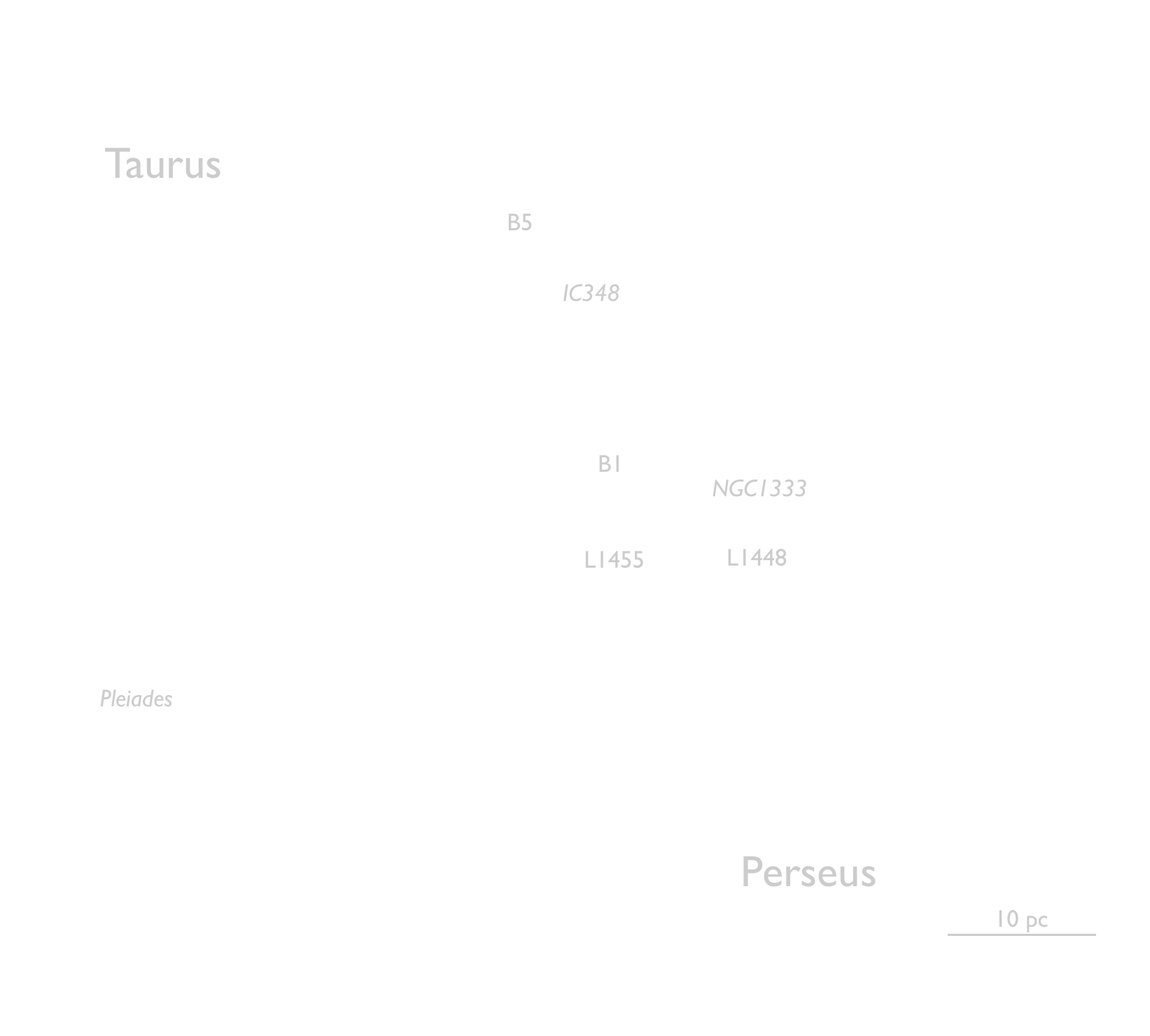}%
  \end{ocg}%

\caption{Composite three-colors image showing \emph{Herschel}/SPIRE intensities for the region considered, where available (with the \SIlist{250;350;500}{\um} bands shown in blue, green, and red). For regions outside the Herschel coverage, we used the \emph{Planck}/IRAS dust model ($\tau_\mathrm{850}$, T, $\mathrm{\beta}$) to
predict the intensity that would be observed at the SPIRE passbands.
\ToggleLayer{fig:2b,fig:2a}{\protect\cdbox{Toggle labels}}}
\label{fig:fig02}
\end{figure*}

\begin{figure*}
\includegraphics[width=\hsize]{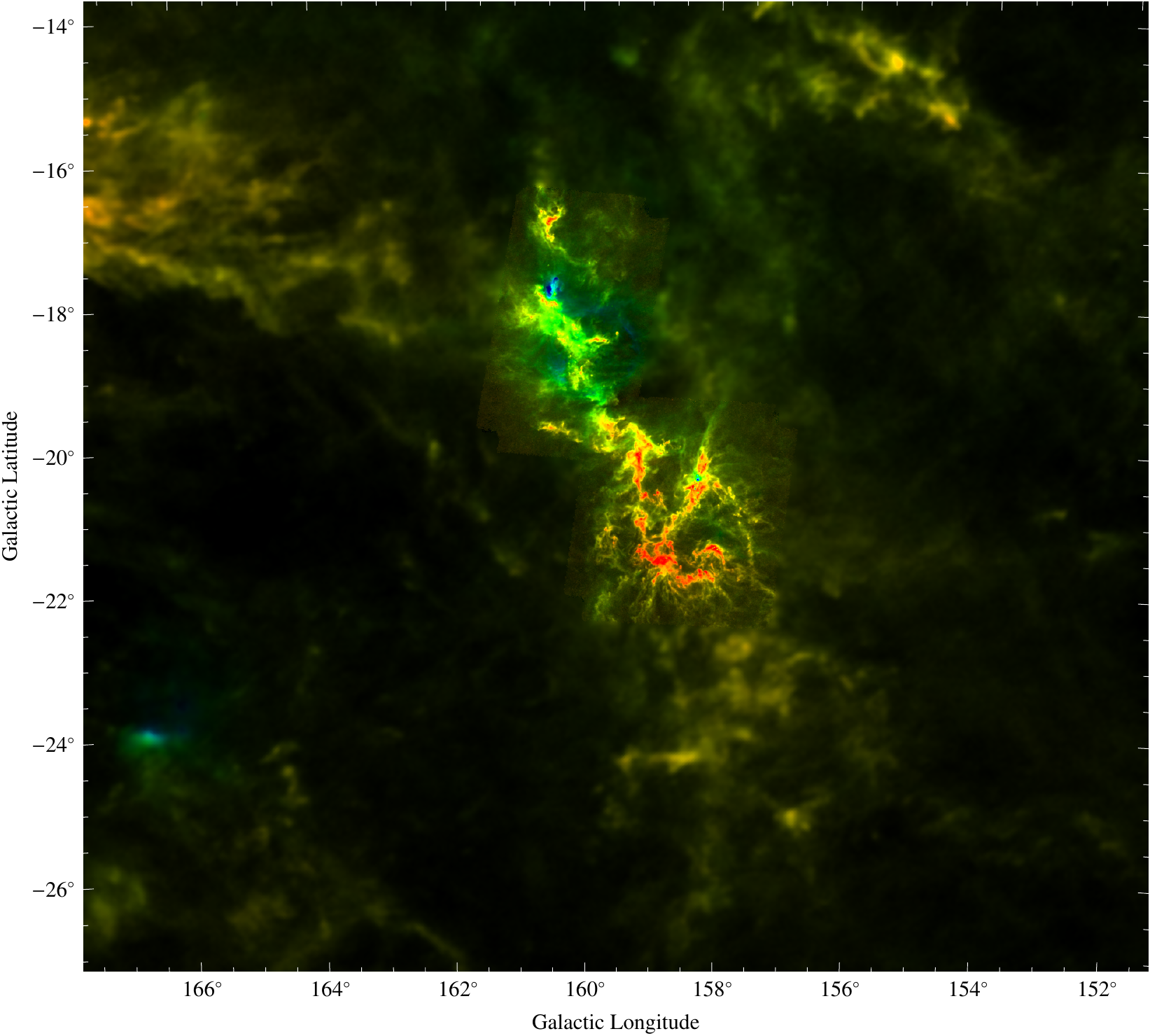}
\caption{Combined optical depth-temperature map for Perseus. The image shows the optical depth as intensity and the temperature as hue, with red (blue) corresponding to low (high) temperatures.}
\label{fig:fig03}
\end{figure*}

\subsection{SED fitting}
The aim of our work is to infer the effective dust color temperature and the optical depth from the fluxes measured by \emph{Herschel} at \SIlist{250; 350;500}{\um}, following \citet{Lombardi2014}, and also PACS \SI{160}{\um} where available.  For this purpose, we first convolved all the \emph{Herschel} data to the beam size of SPIRE \SI{500}{\um}, i.e. to $36\, \mathrm{arsec}$ and then we fitted the observed spectral energy distribution with the function \eqref{eq:emission}, that was integrated over the \emph{Herschel} bandpasses. 
The detailed fitting procedure is described in detail in \cite{Lombardi2014}, however we report here a brief outline of the basic steps used to generate the maps:
\begin{itemize}
\item For each SPIRE band we multiply for the correcting factor $C = K_{4e}/K_{4p}$. Since the extended source calibration method changed as a consequence of the work by \cite{Griffin2013}, the correcting factors  are not the same as those used in \cite{Lombardi2014}. Specifically, $C$ is now equal to $(0.9986, 1.0015, 0.9993)$, while before it was $C = (0.9828, 0.9834, 0.9710)$ for the \SIlist{250;350;500}{\um} bands respectively. However, this change doesn't have a significant impact on the final maps. 
\item We perform an absolute flux calibration for the \emph{Herschel} bands, using \emph{Planck} maps.
\item We assume a modified black body SED and we compute the expected flux at each reference passband.
\item We modify the SED until we obtain a good match between the observed and theoretical fluxes. For this step we use a simple $\chi^2$ minimization technique that takes into account the calibration errors.
\end{itemize}

For the analysis presented here, we used the HIPE \citep{2010ASPC..434..139O} extended emission level 2.5 products for SPIRE data and Unimaps for PACS data. For PACS data it is possible to use Scanamorphos, however there is no significant difference between the two maps. 

Figure \ref{fig:fig02} shows a color-composite image of the combined reduced \emph{Herschel}/SPIRE data for the region considered here, together with the predicted fluxes from Planck at the three SPIRE passbands for locations outside the \emph{Herschel} coverage.

\subsection{Optical depth and temperature maps}

\begin{figure}
  \centering
  \begin{ocg}{fig:3a}{fig:3a}{0}%
    \includegraphics[width=\hsize]{fig03b}%
  \end{ocg}%
  \hspace{-\hsize}%
  \begin{ocg}{fig:3b}{fig:3b}{1}%
    {\color{white}\rule{\hsize}{0.63\hsize}}%
    \hspace{-\hsize}%
    \includegraphics[width=\hsize]{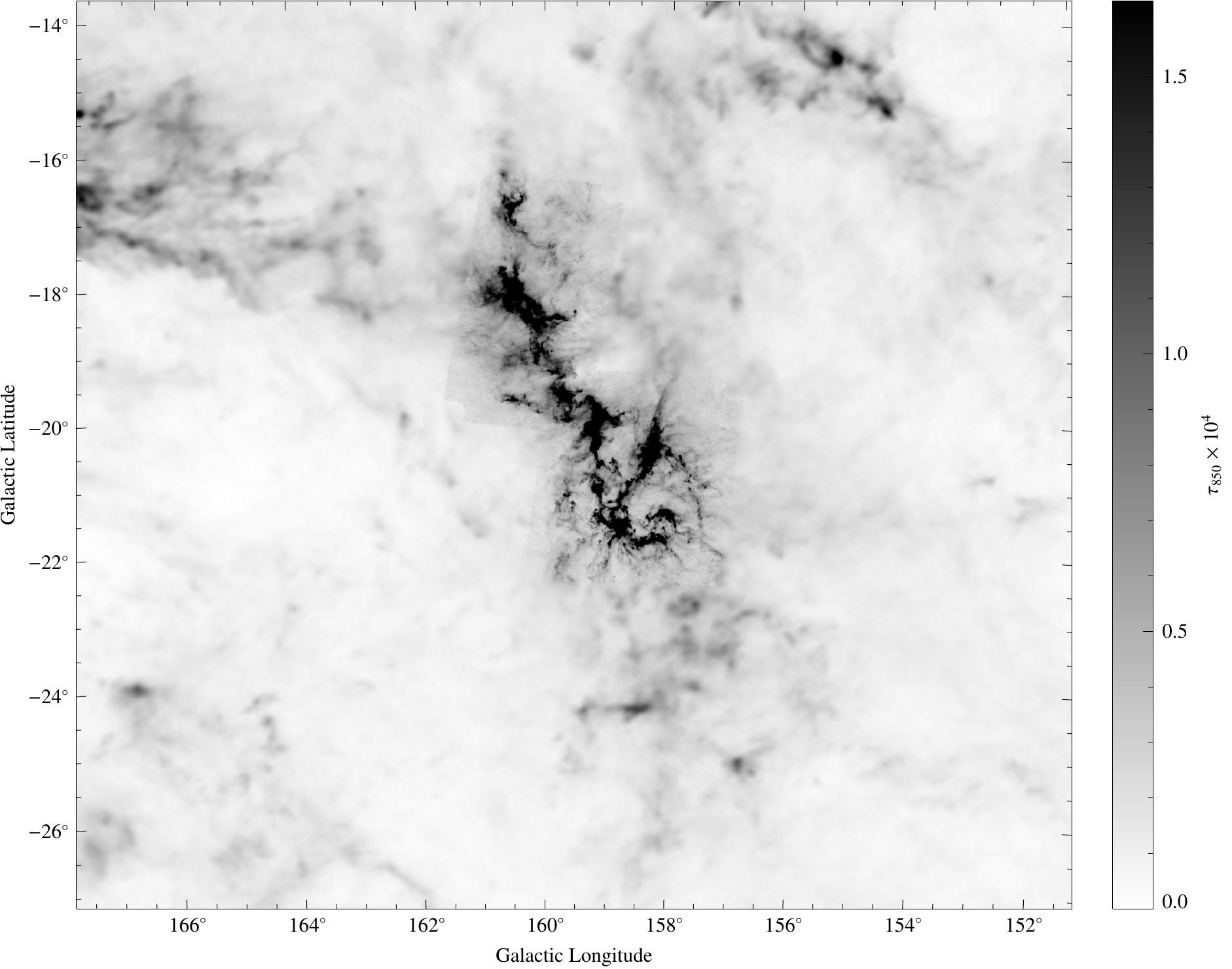}%
  \end{ocg}%
\caption{Optical depth  map of the field and, on a \ToggleLayer{fig:3a,fig:3b}{\protect\cdbox{different layer}}, the corresponding error map. This figure, as the following one (Fig. \ref{fig:opac}), is produced by the method  described in the text, using the reduced fluxes of SPIRE and the expected fluxes at SPIRE frequencies deduced from the \emph{Planck} maps \citep{Planck2014}. \label{fig:opac}}
\end{figure}

\begin{figure}
  \centering
  \begin{ocg}{fig:4a}{fig:4a}{0}%
    \includegraphics[width=\hsize]{fig03d}%
  \end{ocg}%
  \hspace{-\hsize}%
  \begin{ocg}{fig:4b}{fig:4b}{1}%
    {\color{white}\rule{\hsize}{0.63\hsize}}%
    \hspace{-\hsize}%
    \includegraphics[width=\hsize]{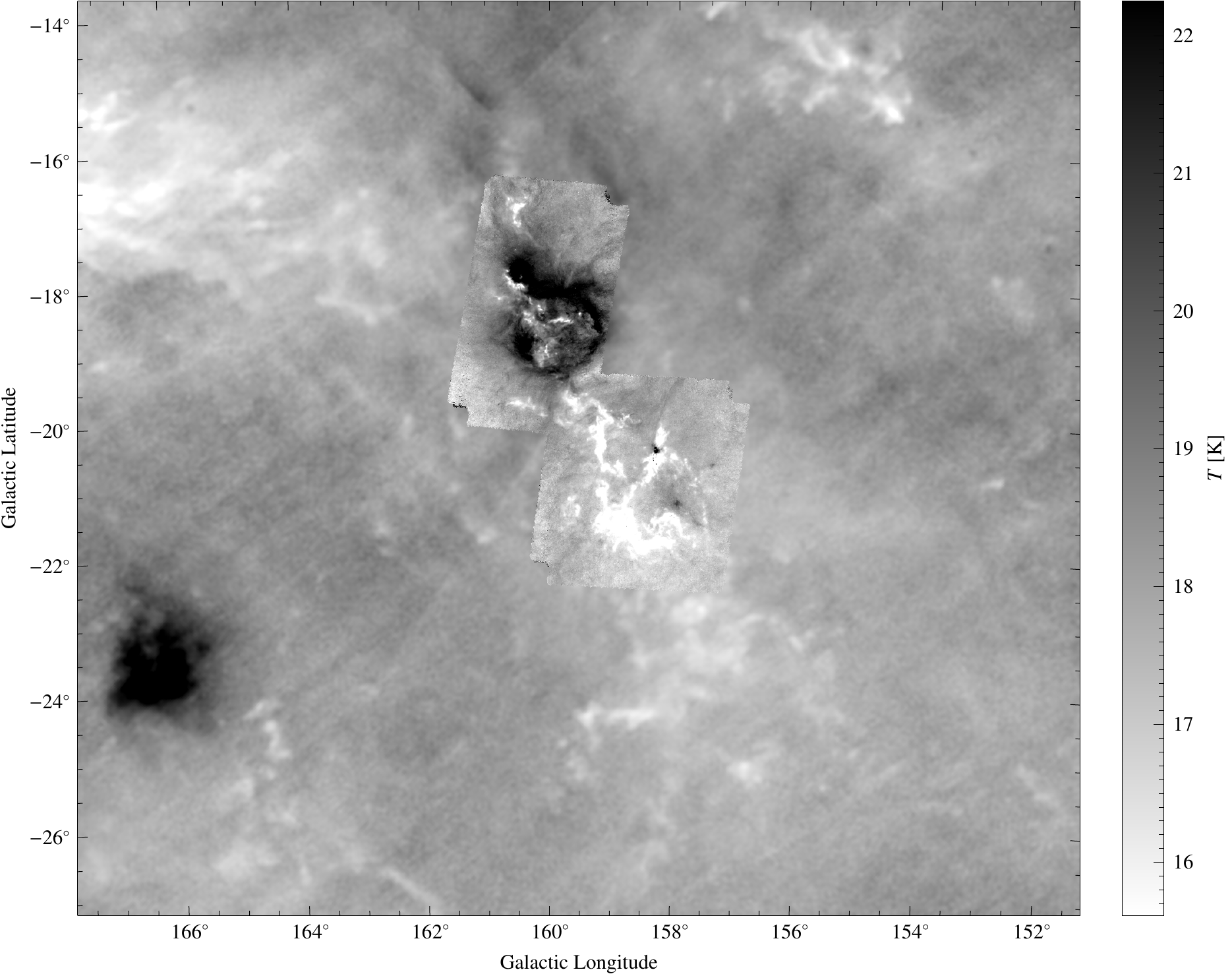}%
  \end{ocg}%
\caption{Temperature map of the field, and, on a \ToggleLayer{fig:4a,fig:4b}{\protect\cdbox{different layer}}, the corresponding error map.\label{fig:temp}}
\end{figure}

Figure \ref{fig:fig03} shows the combined optical depth-temperature map: the effective dust temperature is represented using different values of hue, while the intensity is proportional to the optical depth. Figures \ref{fig:opac} and \ref{fig:temp} individually show the temperature and optical depth maps.

Note that \emph{Planck} data have lower resolution (5 arcmin instead of 36 arcsec) and a significantly lower noise.
In particular, the error on the regions covered by \emph{Planck}  data is much smaller than that on the \emph{Herschel} areas.  Figs. \ref{fig:opac} and \ref{fig:temp} highlight the differences in optical depth and temperature that characterize the cloud.

Referring to the temperature map (Fig.~\ref{fig:temp}), we can note that the regions IC 348 and G159.6-18.5 present a higher temperature than the rest of the cloud and  the ring of dust that surround the star HD 278942 \citep{Andersson2000} is particularly visible. The temperature ranges between $\sim 10 \, \mathrm{K}$ and $\sim 36 \, \mathrm{K}$. 

\subsection{Extinction conversion}
\begin{figure}
\includegraphics[width=\hsize]{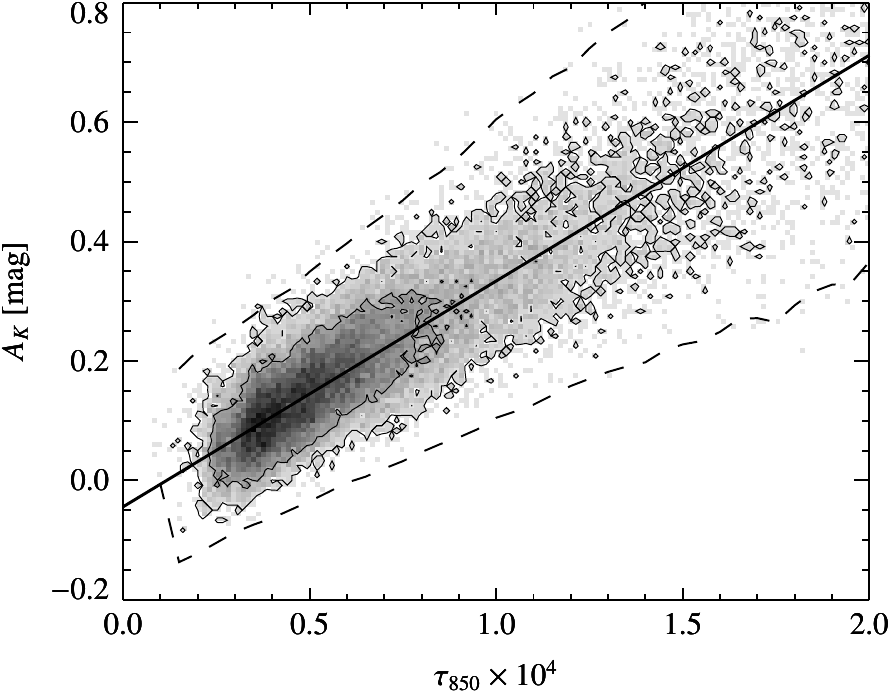}
\caption{Relationship between submillimiter optical-depth and NIR extinction in Perseus. The image shows the best linear fit, used to calibrate the data (solid line), together with the expected $3\sigma$ region (dashed lines), as calculated from direct error propagation in the extinction map. The contours enclose respectively the 68\% and the 95\% of the points.}  \label{fig:fit_lin}
\end{figure}

In order to compare the results obtained here with other observations, we converted the optical depth $\tau_{850}$ to the extinction $A_K$. 
For this purpose, we smoothed the optical depth map to the same resolution as the extinction map (2.5 arcmin) of the region \citep{Lombardi2010}. We assumed that within the range $0 - 2\times10^{-4}$, the law that describes the relation between $\tau_{850}$ and $A_{\it{K}}$ is linear (see Fig. \ref{fig:fit_lin}):
\begin{equation}\label{eq:linear-conversion}
A_{\it{K}} = \gamma\tau_{850}+ \delta.
\end{equation}
The slope $\gamma$ is proportional to the opacity $\kappa_{850}$ and to the extinction 
coefficient in the $K$ band, $C_{\mathrm{2.2}}$. Since the extinction in $K$ band is defined as:
\begin{equation}
A_K = -2.5\log_{10}\left(\frac{I_{\mathrm{obs}}}{I_{\mathrm{I_{true}}}}\right) = 2.5(\log_{10}\mathrm{e})\,C_{2.2}\Sigma_\mathrm{dust},
\end{equation}
the relationship between $\gamma$ and $C_{2.2}$ is simply $\gamma~\simeq~1.0857~C_{2.2}~/~\kappa_{850}$.
The coefficient $\delta$ is associated with uncertainties in the absolute flux calibration of Herschel, or to uncertainties in the extinction maps, or both. Through a $\chi^2$ minimization method,  we obtained the following values for the two parameters:
\begin{equation}
\left\{
\begin{aligned}
\gamma & =  3931\pm 274 \,\mathrm{mag}, \\
\delta & = -0.05 \pm -0.02 \times 10^{-6} \, \mathrm{mag}.\\ 
\end{aligned}
\right.
\end{equation}
The fact that $\delta$ is close to zero confirms the goodness of the calibration of \emph{Herschel} data and of the extinction map.
The value obtained for the slope $\gamma$ is quite close to the ones found by \cite{Lombardi2014} for Orion A and B ($\gamma_{\mathrm{Orion\,A}}=2640 \, \mathrm{mag}$ and $\gamma_{\mathrm{Orion\,B}}=3460 \, \mathrm{mag}$). Being proportional to the extinction coefficient $C_{2.2}$ and the opacity at $\SI{850}{\um}$, the coefficient $\gamma$ is related to the dust composition and grain distribution, and therefore differences in the values of $\gamma$ are likely related to differences in these quantities.

\begin{figure}
\begin{center}
\includegraphics[width=\hsize]{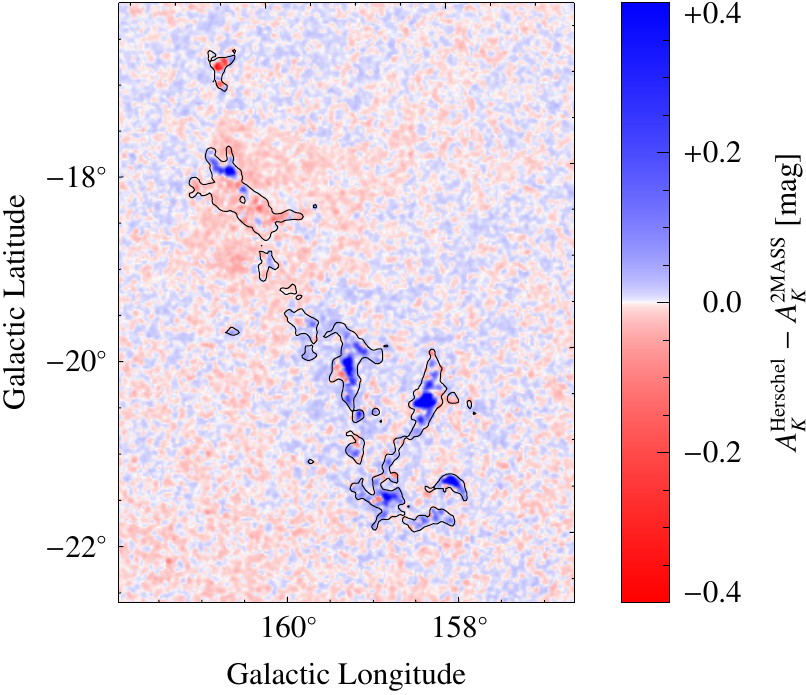}
\end{center}
\caption{Difference between the extinction values evaluated with the 2MASS/\textsc{NICEST} map and the \emph{Herschel} opacity map. The contour represents the $A_K = 0.5\,\mathrm{mag}$ level.\label{fig:diff}}
\end{figure}

If we consider a wider range of values (see Fig. \ref{fig:05b-nicest}), with $\tau \times 10^4 \le 10$, the relation unexpectedly deviates from the linear law. In this case the law between $\tau_{850}$ and $A_{\it{K}}$ can be fitted by the empirical relation:
\begin{equation}\label{eq:power_law}
A_{\it{K}} = c_1 + c_2\tau_{850}^{c_3}.
\end{equation}
The best fit values for the three parameters are:
\begin{equation}
\left\{
\begin{aligned}
c_1 &= -0.38447 \pm   6.9 \times 10^{-05} \,\mathrm{mag},\\
c_2 &=  61 \pm 5  \,\mathrm{mag} ,\\
c_3 &= 0.47973\pm 2.6  \times 10^{-05}.\\
\end{aligned}
\right.
\end{equation}
The deviation from the linear regime might be due to either the inclusion of regions in the extinction map where a large number of embedded sources are present or the lack of background stars at highest extinction. In order to better understand the problem, we created a map (Fig. \ref{fig:diff}) that shows the difference between the extinction values evaluated from the extinction map and from the \emph{Herschel} data using the linear scaling of Eq.~\eqref{eq:linear-conversion}. The highest discrepancies roughly coincide with the position of the Class 0 sources reported by \cite{Sadavoy2014}. Since in these regions the difference is positive, it seems that the extinction map produced through 2MASS data with the \textsc{NICEST} method underestimates the extinction values in the densest regions, where the number of background stars is low and contamination from embedded sources may be present. For this reason we will consider only the linear case Eq.~ \eqref{eq:linear-conversion} to convert optical depth values to extinction for our \emph{Herschel} optical depths.

\section{Results \label{Results}}

In order to characterize the maps, we produced the integral area function of the cloud, $S(>A_{\it{K}})$, i.e. the surface of the cloud above a certain extinction threshold as a function of that threshold. \cite{Lada2013} observed that the shape of the function $S(>A_{\it{K}})$ and of its derivative, $S'(>A_{\it{K}})$, influences the variation in the rate of star formation of the cloud, especially in the high extinction regions. 
\begin{figure}
\includegraphics[width=\hsize]{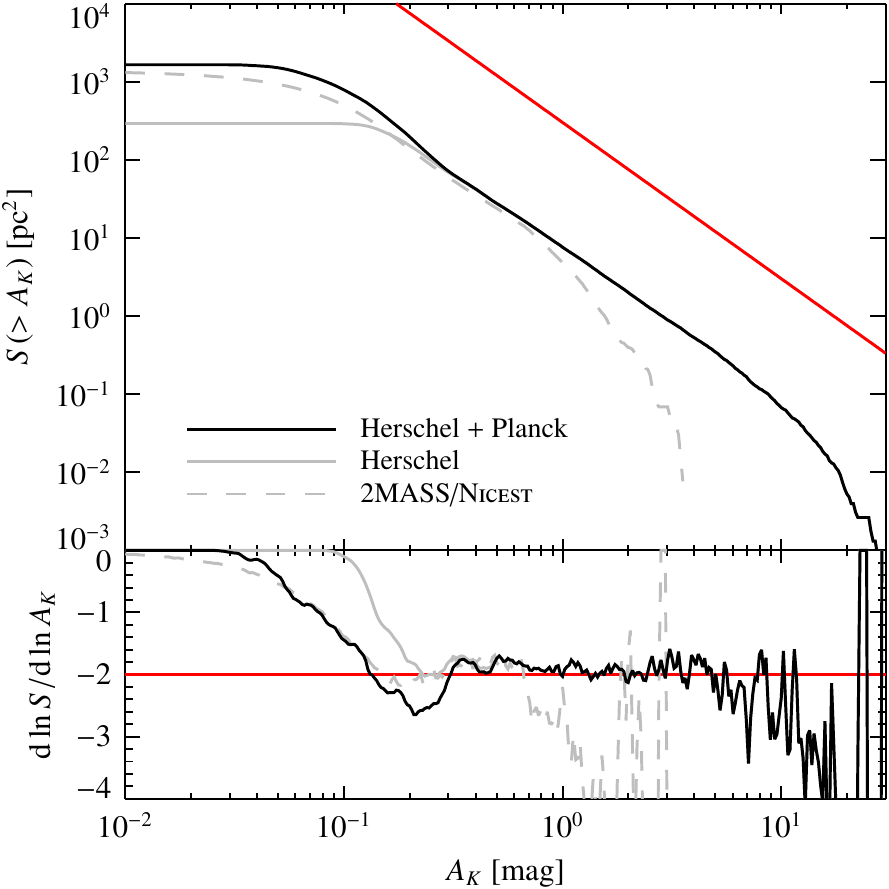}
\caption{Integral area-extinction relation for Perseus, i.e., the physical cloud area above a given extinction threshold as a function of that threshold. The solid black line shows the result for the entire field, while the solid gray line shows the same quantity for the \emph{Herschel} regions. The dashed line shows again the same quantity, but for the  2MASS/\textsc{NICEST} data. The red line shows the slope of the power law $S(>A_{K}) \propto A_{K}^{-2}$. \label{fig:S(Ak)}}
\end{figure}

\begin{figure}
\includegraphics[width=\hsize]{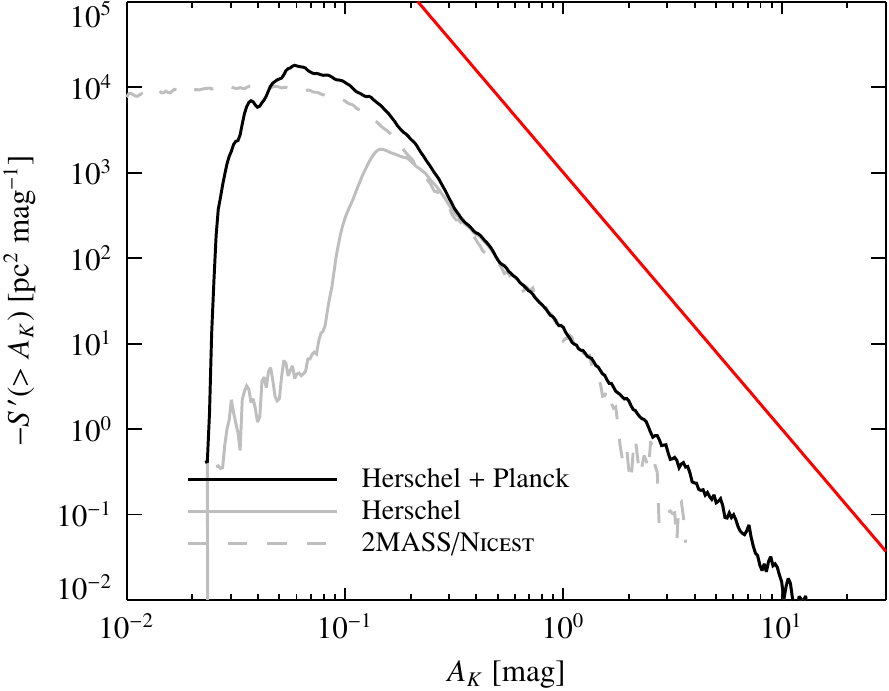}
\caption{Function $ - S'(>A_{\it{K}})$, i.e. the probability distribution function of the measured column density for Perseus. In this plot a lognormal distribution would appear as a parabola, and a power law as a straight line. The red line shows the slope of the power law $S'(>A_{K}) \propto A_{K}^{-3}$. \label{fig:derivS(Ak)}}
\end{figure}

The plot of $S\left(>A_{\it{K}}\right)$ is shown in Fig. \ref{fig:S(Ak)} for different extinction measurements (\emph{Herschel}, \emph{Herschel} and \emph{Planck}, 2MASS/\textsc{NICEST}).  In the same figure, a red line with slope $-2$ is represented. 
Operationally, this plot is built by just counting the sky area above different values of $A_K$.  As such, the specific spacing used for the threshold $A_K$ values is irrelevant (Fig.~\ref{fig:S(Ak)} we used a log-scale in $A_K$).
The law $S(>A_{\it{K}}) \propto A_{\it{K}}^{-2}$ appears to be an excellent description of the area function over two order of magnitudes, from $ A_K \sim 0.1 $ to $A_K \sim 10 \, \mathrm{mag}$. The apparent break of this law at $A_K \sim \SI{15}{mag}$ might either be genuine or due to systematic effects in the \emph{Herschel} maps (unresolved structures, large temperature gradients, flux contamination from point sources). At the other extreme the break at $A_K < \SI{0.15}{mag}$ can be due to various effects, including contamination by unrelated foreground and background material, and inappropriate definition of the cloud boundaries. In reality, as shown by \citet{Lombardi2015}, fundamental constraints, such as contamination by unrelated foreground and background dust emission, severely limit our ability to measure the area function below $A_K < \SI{0.15}{mag}$.

Fig. \ref{fig:derivS(Ak)} shows $-S'(>A_{\it{K}})$, as a function of $A_{\it{K}}$, where $A_{\it{K}}$ is evaluated through Eq. \eqref{eq:linear-conversion}.  This function is proportional to the probability distribution function (PDF) of the column density map and follows a power law with index equal to $-3$.
Note that since we directly derived the function $S(>A_K)$ using a 3-point Lagrangian interpolation (where the spacing $A_K$ is taken into account), this operation effectively produces a result that is proportional to a linearly-binned PDF.  We stress that this differs from the choice adopted by \citet{Lombardi2015}, where PDFs were logarithmically binned. Consequently, the slopes evaluated in this way differ by -1 to the one presented by \citet{Lombardi2015}, but the results are equivalent.

\begin{figure}
\includegraphics[width=\hsize]{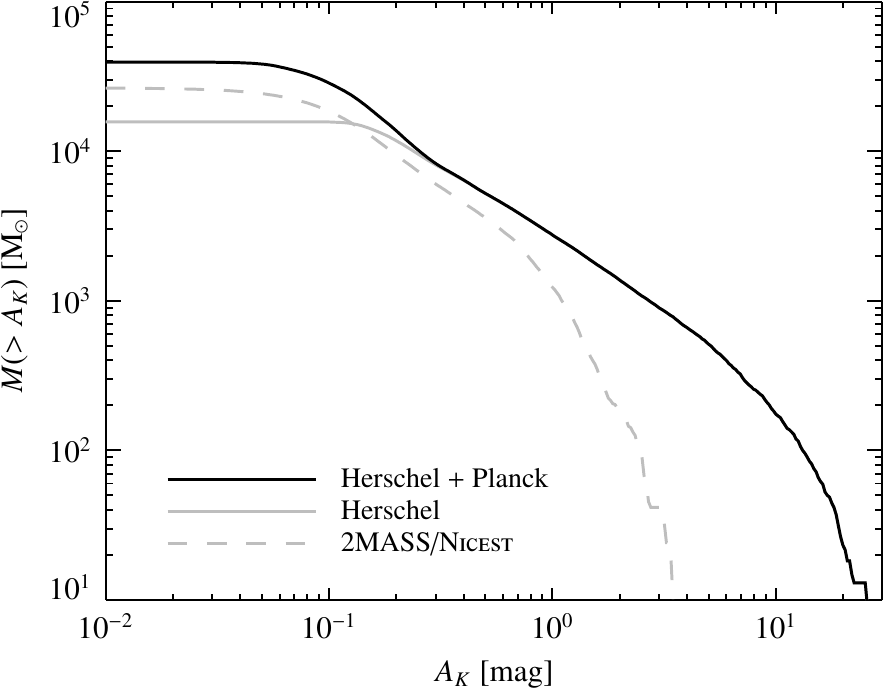}
\caption{Integral mass-extinction relation, i.e. the cloud
mass above a given extinction threshold.  The line codes follow the same convention as in Fig.~\ref{fig:S(Ak)}. The values of the mass above a certain extinction threshold are listed in Table \ref{table:Mass} \label{fig:M(Ak)}}
\end{figure}

\begin{figure}
\includegraphics[width=\hsize]{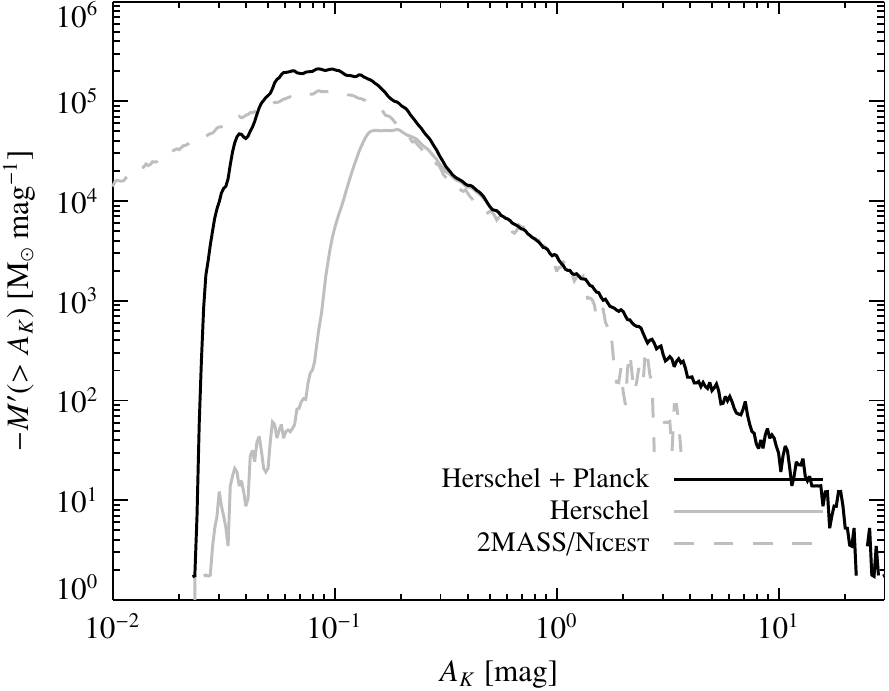}
\caption{Function $-M'(>A_{\it{K}})$, i.e. the derivative of the integral mass function. \label{fig:derivM(Ak)}}
\end{figure}

\begin{table}
\caption{Mass values above an extinction threshold, normalized at the mass at $0.1 \,\mathrm{mag}$, $M_{0.1}$.}
\begin{centering}
\begin{tabular}{ccc}
 $A_{K} [\mathrm{mag}]$ & $M\,[\mathrm{M_{\odot}}]$ & $M/M_{0.1}$\tabularnewline
\hline
\tabularnewline
0.01 &  22153   &  168 \%\tabularnewline
0.1   &    13189 &  100 \%\tabularnewline
0.2	& 7864 &	60 \%\tabularnewline
0.4	& 4956 &	37 \%\tabularnewline
0.6   &   3685 &  27 \%\tabularnewline
0.8   &    2887 &  21 \% \tabularnewline
1.6   &    1532 &  11 \%\tabularnewline
3.2   &    777 &  5 \%\tabularnewline
6.4    &   341 &	2 \% 
\end{tabular}
\par\end{centering}
\label{table:Mass}
\end{table}

Figs. \ref{fig:M(Ak)} and \ref{fig:derivM(Ak)} show the integral function of the mass $M(>A_{\it{K}})$ as a function of $A_{\it{K}}$ and its derivative $-M'(>A_{\it{K}})$. The gas surface density, $\Sigma_{\mathrm{gas}}$, can be expressed as a  mass surface density through the following relation:
\begin{equation}\label{Sigma_Ak_conversion}
\frac{\Sigma_{\mathrm{gas}}}{A_{\it{K}}} = \mu m_p \beta_{\it{K}}\simeq \SI{183}{\Msun}\, \mathrm{{pc}^{-2}} \, \mathrm{mag^{-1}},
\end{equation}
where $\mu = 1.37$ is the mean molecular weight, corrected for the helium abundance, $m_p = 1.67\times10^{-24} \mathrm{g} $ is the proton mass and $\beta_{\it{K}} = 1.67 \times 10^{22} \mathrm{cm^{-2} mag^{-1}} = \left[2N\left(H_2\right)+N\left(H_I\right)\right]/A_{\it{K}}$ is the gas-to-dust ratio. 
The agreement between the solid line (\emph{Herschel} and \emph{Planck} data) and the dashed line (2MASS/\textsc{NICEST}) in Fig.~\ref{fig:S(Ak)} and Fig.~\ref{fig:M(Ak)} is good until $A_{\it{K}}\simeq 0.1\,\mathrm{mag}$, while the discrepancy we observe at high extinction values is due to the higher levels of extinction reached by \emph{Herschel}. 
Similarly to \cite{Lombardi2014}, we can define the mass of the cloud as the surface density integrated over the area:
\begin{equation}
M = \int \Sigma_{\mathrm{gas}}\,dS,
\end{equation}
from which:
\begin{equation}\label{eq:mass}
M \left(>A_{\it{K}}\right) \propto \int_{A_{\it{K}}}^{\infty} \frac{dS\left(>A'_{\it{K}}\right)}{dA'_{\it{K}}} A'_{\it{K}} dA'_{\it{K}}.
\end{equation}
Notice how the power law trend observed in Fig.~\ref{fig:M(Ak)} is a direct consequence of the power law observed in Fig.~\ref{fig:S(Ak)}, by replacing $dS(>A'_K)/dA'_K$ with $A_K^{'-3}$ in Eq. \eqref{eq:mass}.

\section{The local Schmidt law}\label{Section:Schmidt_law}
Another possible application of the column density maps is to study the validity of the Schmidt law \citep{Schmidt1959} in the Perseus molecular cloud. 
Half a century ago, Schmidt conjectured that the rate of star formation, $\Sigma_{\mathrm{SFR}}$ depends on the (projected) gas density, $\Sigma_{\mathrm{gas}}$,  and in particular is a simple power law. Schmidt argued that the index of the power law was $\sim 2$.  
Recently,  a series of observational studies (\cite{Lada2014} and references therein) demonstrated, using observations of protostellar objects, that a Schmidt scaling law exists within a sample of nearby molecular clouds and it is typically characterized by an index $\sim 2$.
In what follows we investigate the Schmidt law, using in particular Class I and Class 0 protostars \footnote{In the following we will refer to Class I and 0 objects as \emph{protostars}, and to Class II, Class I and Class 0 objects as \emph{YSOs}.} to estimate the protostar surface density.
Following \cite{Lada2013}, we decided to consider only such sources, since they are likely still at or close to their original birth place.
To estimate the protostar surface density, $\Sigma_{\mathrm{YSO}}$, we used already existing catalogs based on \emph{Spitzer} data and we built a catalog of young stellar objects through the analysis of the WISE satellite data. 
We used the WISE data mainly to enhance and improve the \emph{Spitzer} based catalogs. Indeed, even though WISE resolution (6.1", 6.4", 6.5", and 12.0" in the four bands) is lower than Spitzer, its all-sky coverage allows to include sources detected in areas not surveyed by \emph{Spitzer}.
To make our classification as complete as possible, we also included the new sources detected by  \citet{Sadavoy2014}.

\subsection{Identifying YSO candidates}
The procedure adopted to search for YSO candidates is based on a comparison of the source distribution in several color-color diagrams between the science and a control field. We choose a control field in the region defined by: 
\begin{equation}
137^{\circ} < l < 147^{\circ} \hspace{0.5cm} -25^{\circ}< b < -15^{\circ}.
\end{equation}
We selected this area based on two requirements: it is characterized by low extinction values and the background object distribution is similar to the one of the science field, since it has the same galactic latitude of the science field, which corresponds to the region defined previously for 2MASS data. 
To classify the sources in the science field, we approximately followed the scheme proposed by \citet{Koenig2012}, which in turn is based on the scheme by \citet{Gutermuth2008} and \citet{Gutermuth2009}. 
Koenig analysed several WISE color-color diagrams of \emph{Spitzer} selected YSOs in the Taurus region. Besides, they characterized the contaminating sources by studying the distribution of objects with declination $> 88^{\circ}.22$ (celestial north pole) in the $W_1-W_2$ vs. $W_2-W_3$ color-color diagram. 
By comparing the science and the control field color-color diagrams, one can note the presence in the science field of sources with color excess: these may be reddened background objects or embedded YSOs. 
As \citet{Rebull2011} noted, no color-color diagram can perfectly find all the YSOs and remove all the contaminants: the contamination rate for any color selection is expected to be large and ancillary data from other observations are often needed to choose the most likely candidates. \citet{Koenig2012} estimated a contamination rate for "typical" star forming regions of about 2.4 objects resembling Class~I YSOs, 3.8 objects resembling Class~II YSOs, and  1.8 objects resembling transition discs per square degree. As a result, in our field, which is $\sim 100 \,\mathrm{deg^2}$, the contaminants expected are respectively $\sim 240$, $\sim 380$, and $\sim 180$.

\begin{figure*}[htbp]
\begin{centering}
\includegraphics[width=0.45\textwidth]{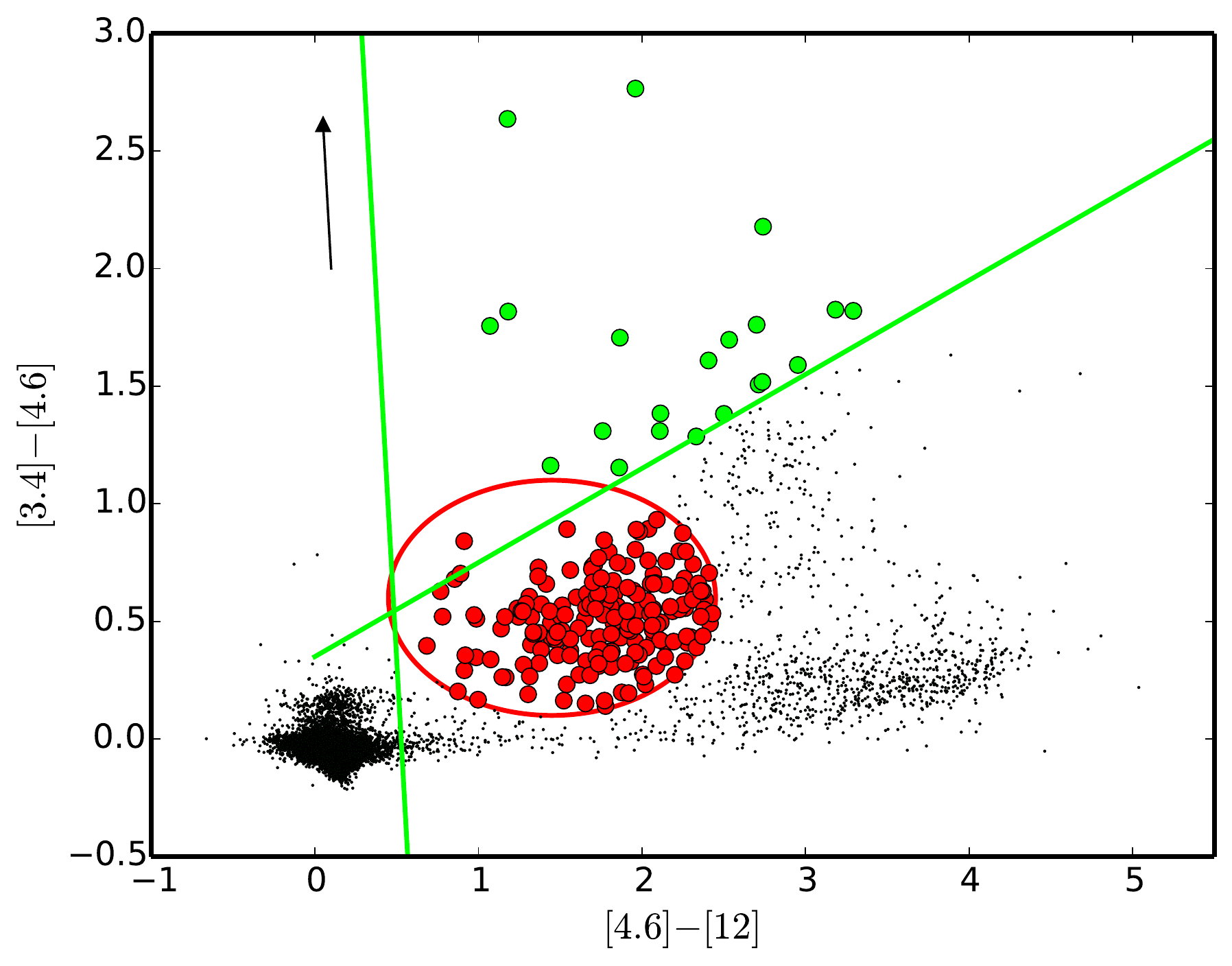}%
\qquad\qquad
\includegraphics[width=0.45\textwidth]{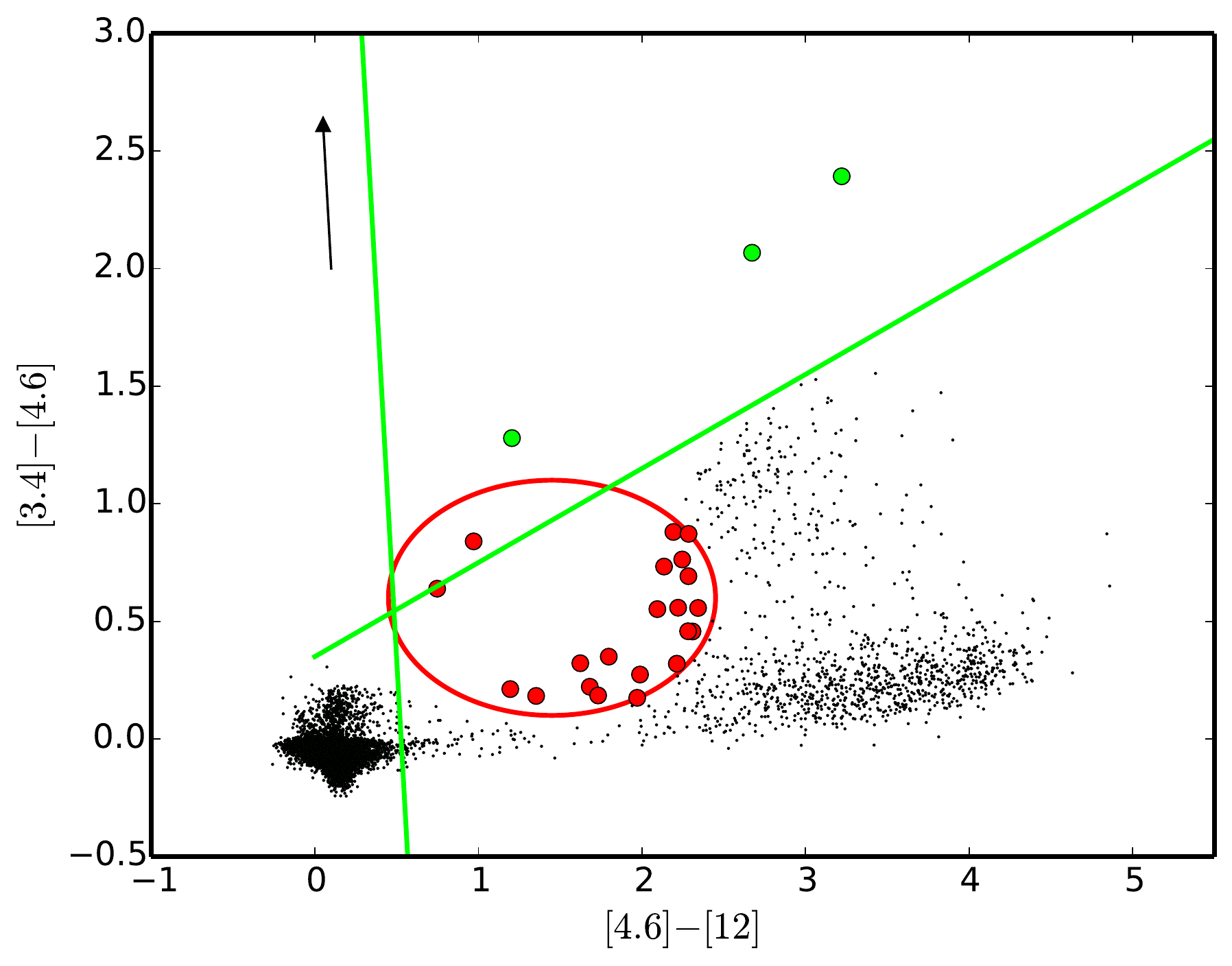}
\caption{$W1 - W2$ ($[3.4]-[4.6]$) vs. $W2 - W3$ ($[4.6]-[12]$) diagram for the science and the control field. Red points represent candidate Class~II objects, green points represent candidate Class~I objects, while black points represent the other sources in the field. The red ellipse and the green lines have equations specified in the text. \label{fig:w123_1}}
\end{centering}
\end{figure*}

\begin{figure*}
\begin{centering}
\includegraphics[width=0.45\textwidth]{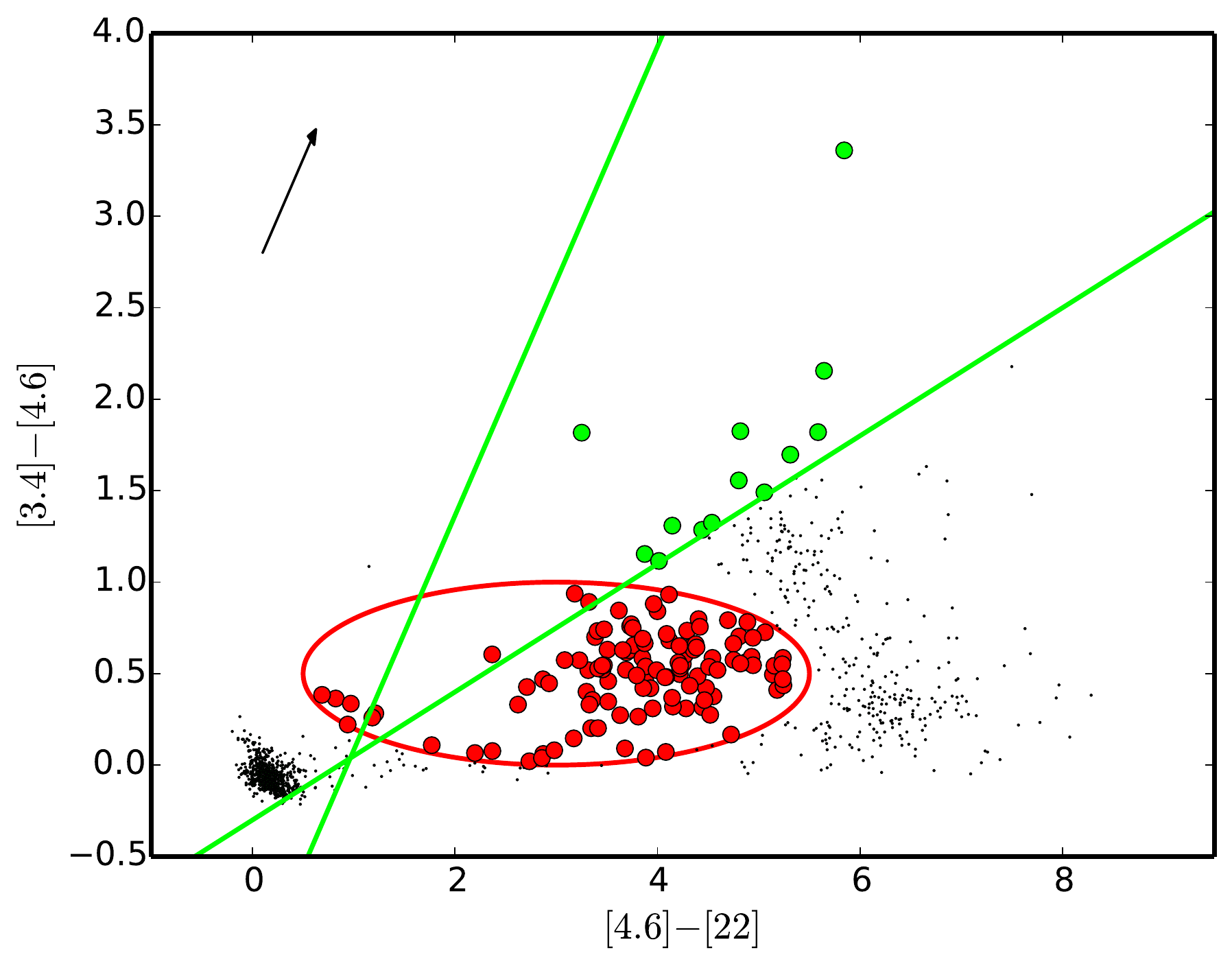}%
\qquad\qquad
\includegraphics[width=0.45\textwidth]{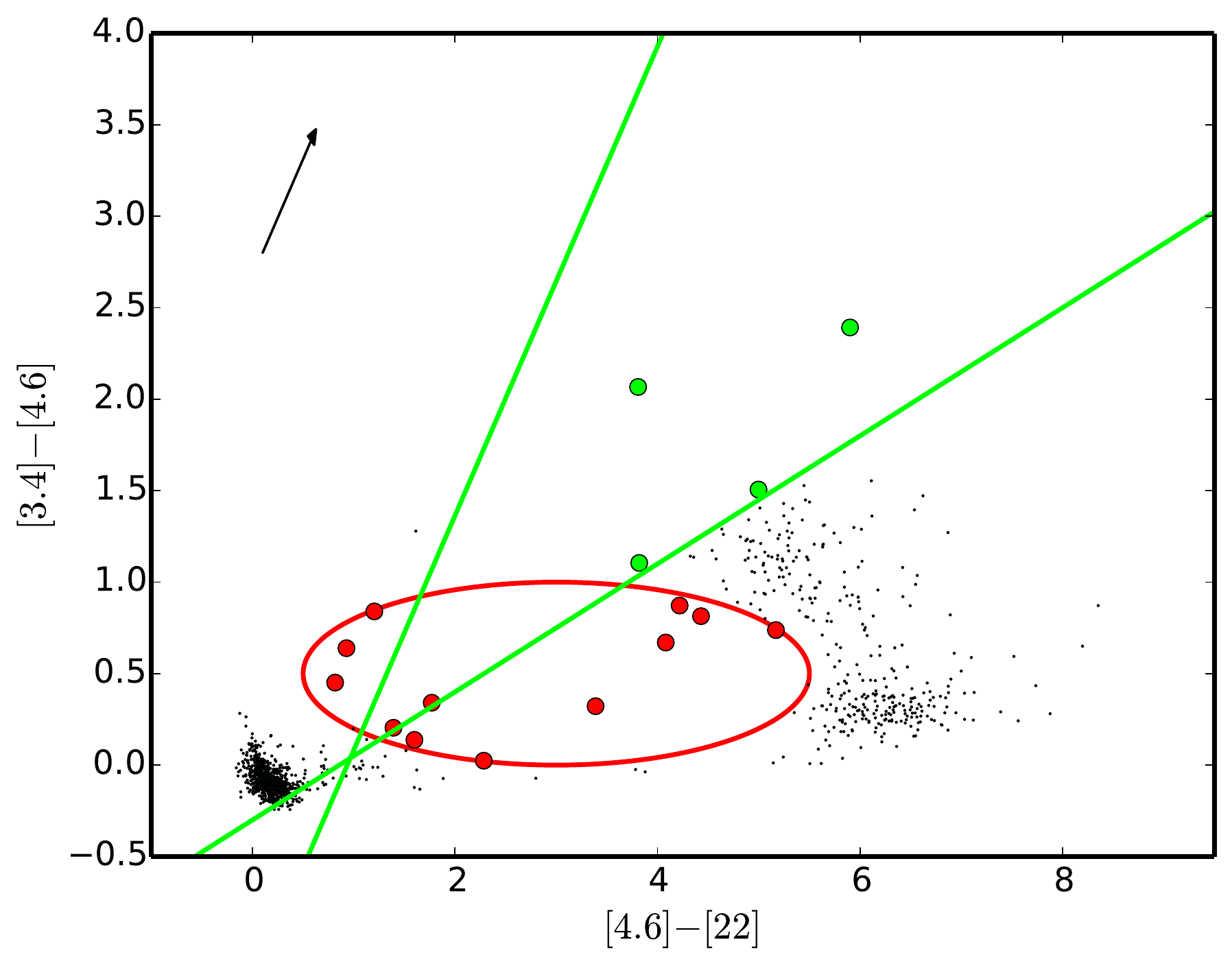}
\caption{$W1 - W2$ vs. $W2 - W4$ diagram of the science and the control field. Red points represent candidate Class~II YSOs, green points represent candidate Class~I protostars, while black points represent the other sources in the field. The red ellipse and the green slopes have equations specified in the text. \label{fig:w124_1}}
\end{centering}
\end{figure*}

Figure \ref{fig:w123_1} shows the color-color diagram in $W_1$, $W_2$ and $W_3$ bands of the science (left) and the control field (right). Comparing them, we observed a significant excess of sources inside the marked red ellipse with center
\begin{equation}
\left\{
\begin{aligned}
 \left[4.6\right]-\left[12\right] &= 1.45 \,\mathrm{mag},\\
 \left[3.4\right]-\left[4.6\right] &= 0.6 \,\mathrm{mag}
\end{aligned}
\right.
\end{equation}
and axes $a=1\, \mathrm{mag}$ and $b=0.5\,\mathrm{mag}$, and between the two blue slopes with equations:
\begin{equation}\label{Eq:w123_slopes}
\left\{
\begin{aligned}
 \left[3.4\right]-\left[4.6\right] &= 0.4\times\left([4.6]-[12]\right) + 0.35  \,\mathrm{mag},\\
 \left[3.4\right]-\left[4.6\right] &= -12.4\times\left([4.6]-[22]\right) + 6.3 \, \mathrm{mag}.
\end{aligned}
\right.
\end{equation}
Therefore, we primarily select 213 objects inside these two regions. The reddening  of the objects inside the ellipse is low, and therefore we can suppose that they are Class~II YSOs. Instead, the objects included between the two slopes (22) are highly reddened and can be classified as Class~I protostars. We can observe that the selected Class~I protostars satisfy Koenig requirements. Besides, our classification is stricter, since we  remove all sources in the AGN/galaxy region of the diagram. Also the selection criteria of Class~II YSOs are close to the ones chosen by \citet{Koenig2012}. Indeed, the minimum color values accepted to include a source in the selection are:
\begin{equation}
\left\{
\begin{aligned}
 \left[4.6\right]-\left[12\right] &= 0.95\, \mathrm{mag},\\
 \left[3.4\right]-\left[4.6\right] &= 0.35\, \mathrm{mag}.
\end{aligned}
\right.
\end{equation}
Figure \ref{fig:w124_1} shows the diagrams $[3.4] - [4.6]$ vs. $[4.6] - [22]$ 
and is analogous to the previous one. Class~II YSOs lie within the red ellipse, with center
\begin{equation}
\left\{
\begin{aligned}
 \left[4.6\right]-\left[22\right] &= 3\, \mathrm{mag},\\
 \left[3.4\right]-\left[4.6\right] &= 0.5\, \mathrm{mag},
\end{aligned}
\right.
\end{equation}
and axes $a=2.5\,\mathrm{mag}$ and $b=0.5\,\mathrm{mag}$, while Class~I protostars are included between the slopes with equations:
\begin{equation}
\left\{
\begin{aligned}
 \left[3.4\right]-\left[4.6\right] &= 0.4\times\left([4.6]-[22]\right) - 0.3  \,\mathrm{mag},\\
 \left[3.4\right]-\left[4.6\right] &= 1.3\times\left([4.6]-[22]\right) - 1.2 \, \mathrm{mag}.
\end{aligned}
\right.
\end{equation}

The total number of sources is 130: 117 of them are classified as Class~II YSOs, while the remaining 13 as Class~I protostars. Further, we can observe the similarity with Koenig criteria.

\begin{figure}
\includegraphics[width=\hsize, keepaspectratio]{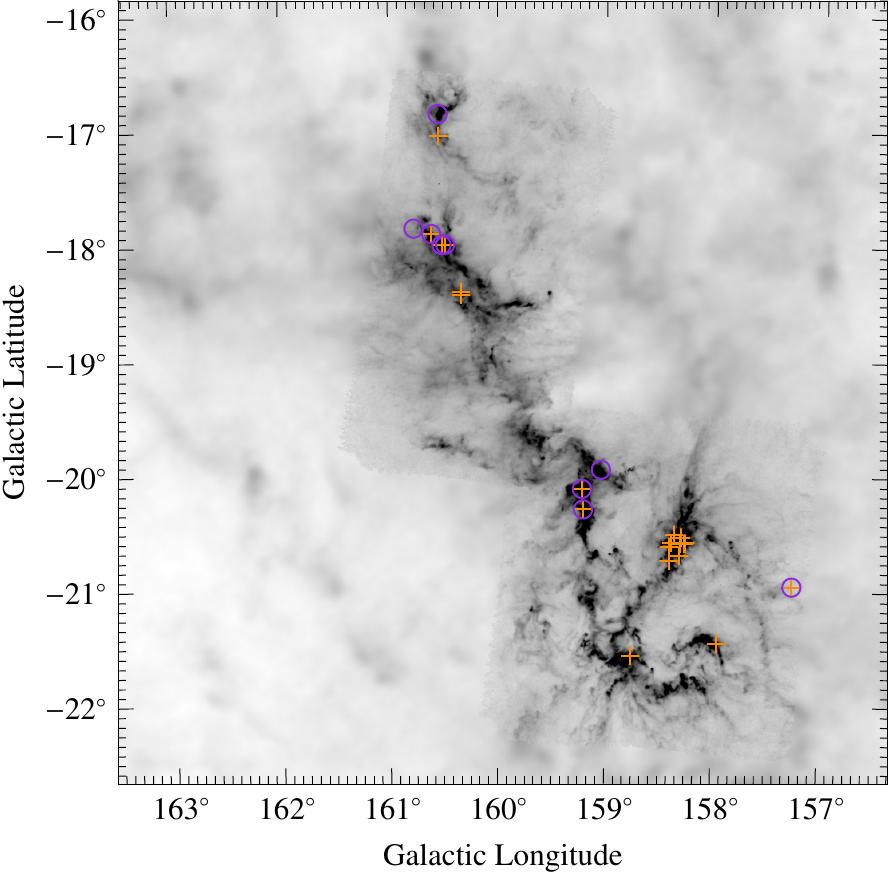}
\caption{Spatial distribution of candidate Class~I objects (orange crosses) selected through the $[3.4]-[4.6]$ vs. $[4.6]-[12]$ color-color diagram and Class~I objects (purple circles) selected through the $[3.4]-[4.6]$ vs. $[4.6]-[22]$ color-color diagram. The map shown here is the optical depth map presented before. \label{fig:w123_ClassI}}
\end{figure}

\begin{figure}
\includegraphics[width=\hsize, keepaspectratio]{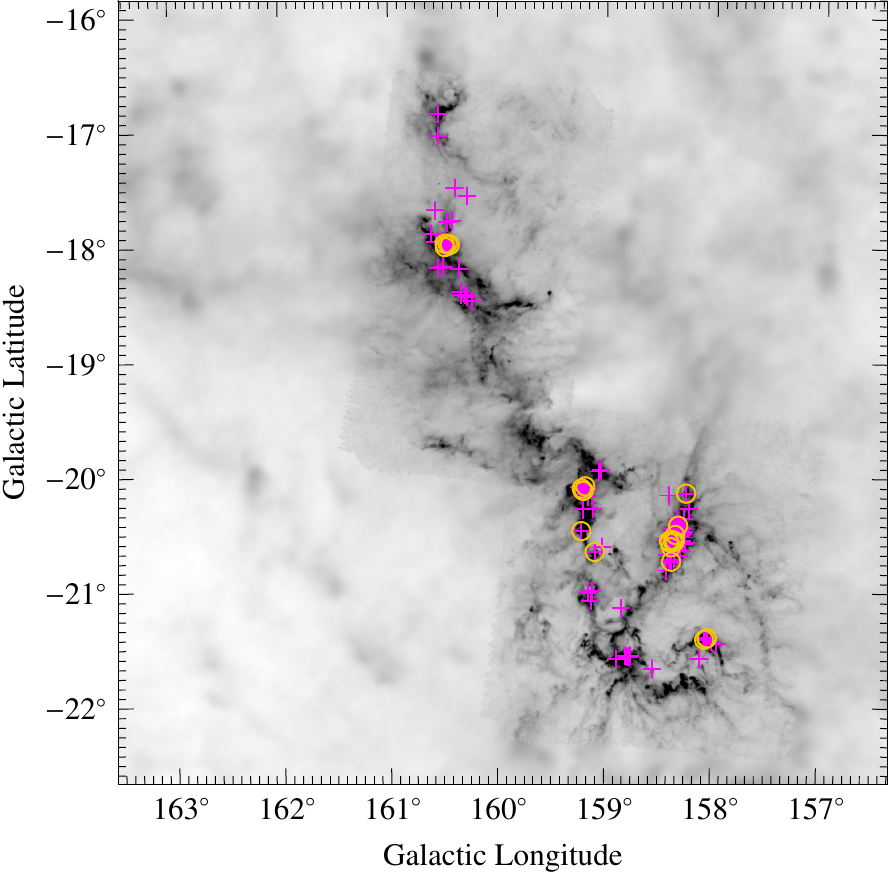}
\caption{Spatial distribution of the candidate Class~I (pink crosses) and Class~0 objects (orange circles, see \citet{Sadavoy2014}) used to estimate the Schmidt law. \label{fig:catalog_I0}}
\end{figure}

To find YSO candidates associated with known objects, we cross-checked our selection with SIMBAD astronomical database and with the catalogs by \citet{Evans2009, Gutermuth2008, Gutermuth2009, Jorgensen2006, KirkMyers2011,Winston2009} and \citet{Young2015}, based on \emph{Spitzer} data. 
Many sources selected through WISE data analysis were already included in other catalogs. Nevertheless, the complete sky coverage of the WISE satellite allows to study regions that are parts of the cloud but were never observed by \emph{Spitzer}. 
The final selection consists of previously identified \emph{Spitzer} sources merged with newly identified WISE objects. 61 new sources were selected through WISE data, of which 17 belong to IC~348 and just one to NGC~1333. 

We now focus on the candidate Class~I protostars selected previously with our color-color diagrams.
To further anlayse these sources, in particular to test whether our classification scheme is reliable, it is useful to study their spatial distribution. Figure \ref{fig:w123_ClassI} shows the positions of the Class~I protostars that were selected through the $[3.4]-[4.6]$ vs. $[4.6]-[12]$ and $[3.4]-[4.6]$ vs. $[4.6]-[22]$ diagrams.
All sources (except one) are placed in the highest opacity regions: this agrees well with the hypothesis that they are located close to their original place of formation. 
Most of Class~I YSOs selected through the $[3.4]-[4.6]$ vs. $[4.6]-[12]$ diagram are situated in the NGC~1333 region. This could imply, as \citet{Jorgensen2006} observed, that the age of NGC~1333 is slightly lower than the age of IC~348: indeed, it is estimated that IC~348 is 2~Myrs old \citep{Muench2007}, while NGC~1333 does not exceed 1~Myr \citep{Lada1996}.
We already noted that the number of sources selected with the $[3.4]-[4.6]$ vs. $[4.6]-[22]$ is lower than for the $[3.4]-[4.6]$ vs. $[4.6]-[12]$ diagram. 
Class~I protostars in $[3.4]-[4.6]$ vs. $[4.6]-[12]$ diagram correspond to Class~I protostars in $[3.4]-[4.6]$ vs. $[4.6]-[22]$ diagram, except for three sources not detected in the first diagram. Surprisingly, the $[3.4]-[4.6]$ vs. $[4.6]-[22]$ diagram does not detect any Class~I protostar in NGC~1333.
We evaluated the cloud temperature and optical depth at each object position. 
We excluded from the WISE selection all the objects in areas with optical depth $\tau<3.141 \times 10^{-5}$ (corresponding to $A_{K}<0.1\,\mathrm{mag}$), supposing that this level of optical depth corresponds to the cloud boundaries. 
Figure \ref{fig:catalog_I0} shows the final selection of Class~I and Class~0 protostars (139 sources). Only 2 of the new sources selected by WISE are classified as Class~I protostars. All the other sources in the figure have been previously classified using \emph{Spitzer} data. 
We noted that for some sources our classification differs from the one obtained by \emph{Spitzer} catalogs. In those cases, we decided to follow the \emph{Spitzer} classification.

\subsection{Determining the Schmidt law}
\begin{figure}
\includegraphics[width=\hsize, keepaspectratio]{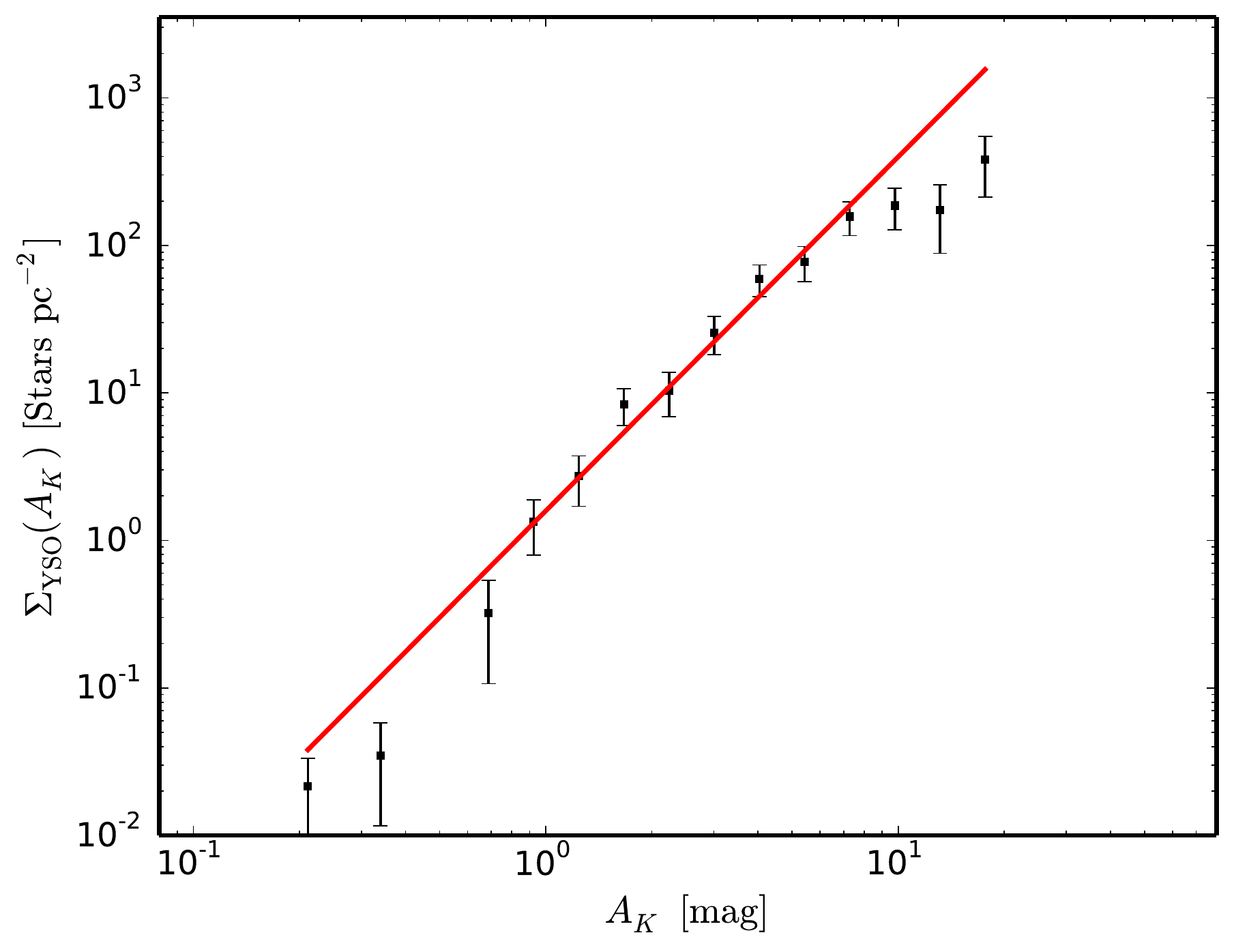}
\caption{The Schmidt law in the Perseus molecular cloud. Here, $\Sigma_{\mathrm{YSO}} \propto A_K^{2.4}$, and $A_K \propto \Sigma_\mathrm{gas}$. \label{fig:Schmidt}}
\end{figure}

\begin{figure}
\includegraphics[width=\hsize, keepaspectratio]{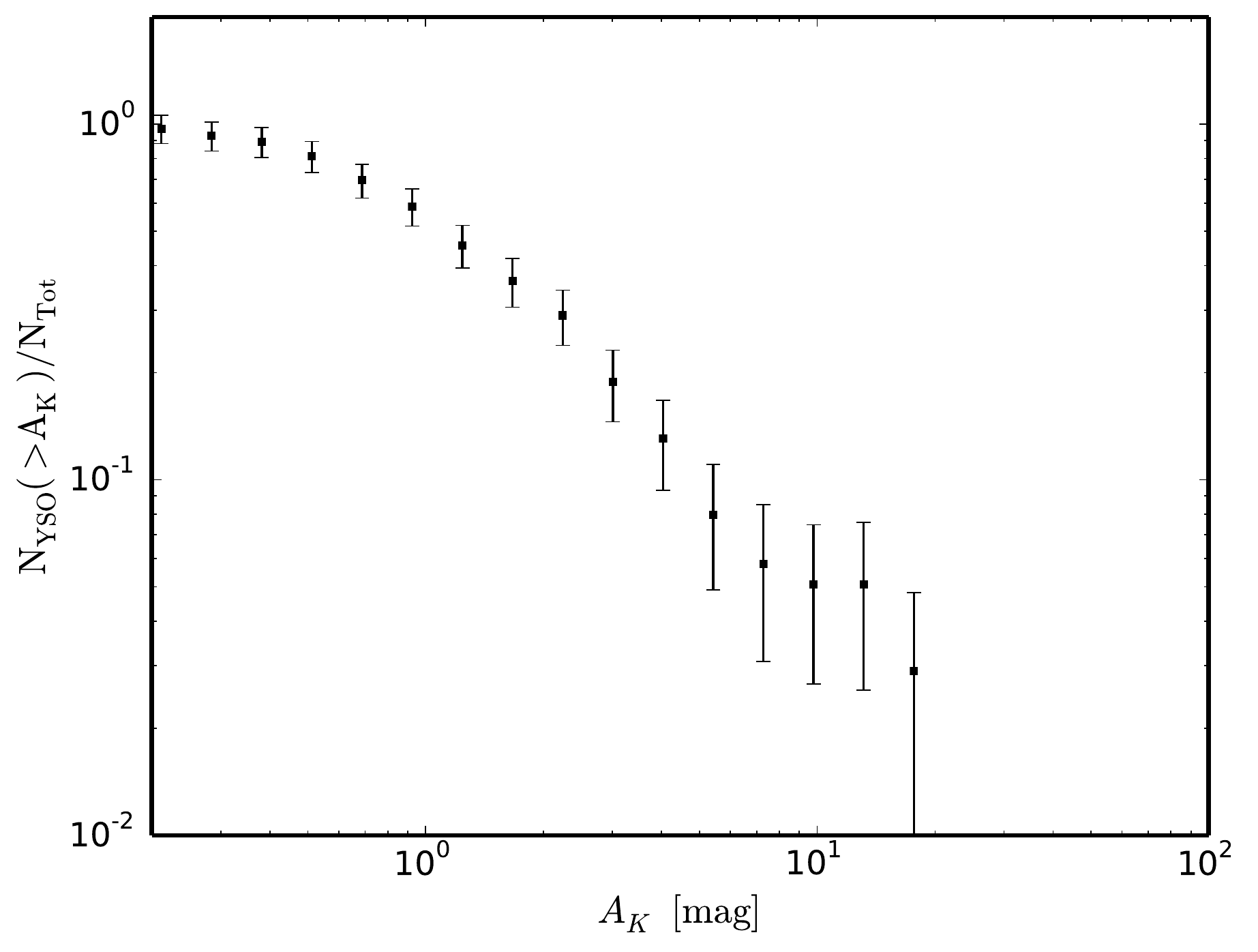}
\caption{The cumulative protostellar function in the Perseus cloud as a function of $A_K$, i.e. $N_\mathrm{YSO}(>A_{K})$. \label{fig:N(Ak)}}
\end{figure}

We computed the extinction at each source position using our \emph{Herschel} derived extinction map and binned them in log-spaced bins.  
Then we evaluated the cloud area between two consecutive extinction levels and the number of sources included within them. 
In this way, we evaluated the protostar surface density, $\Sigma_\mathrm{YSO}$, as a function of the extinction, as the number of sources between two consecutive extinction levels divided by the corresponding area.
Since the gas surface density $\Sigma_{\mathrm{gas}}$ is proportional to the $K$ band extinction (see Eq. \eqref{Sigma_Ak_conversion}), it is possible to express the Schmidt law as:
\begin{equation}
\Sigma_{\mathrm{YSO}}=\kappa A_{K}^{\beta},
\end{equation}
where $\Sigma_{\mathrm{YSO}}$ is simply equal to $\Sigma_{\mathrm{SFR}}\times \tau$, and $\tau$ is the mean age of a Class~I/0 protostar. 
Figure \ref{fig:Schmidt} represents the Schmidt law for the Perseus molecular cloud. The values obtained for the parameters are
\begin{equation}
\left\{
\begin{aligned}
\kappa &= 0.2 \pm 0.07 \, \mathrm{[YSO\, pc^{-2} \, mag^{-\beta}]},\\
\beta &= 2.4 \pm 0.6.
\end{aligned}
\right.
\end{equation}
The value we found for $\beta$ is steeper (although compatible) than what reported by \citet{Lombardi2014} for the Orion molecular cloud complex and by \citet{Lada2013} for Taurus molecular cloud. Indeed, \citet{Lombardi2014} found $\beta= 1.99\pm 0.05$ for Orion A and $\beta = 2.16\pm 0.10$ for Orion B, while \citet{Lada2013} found $\beta = 2.09\pm 0.14$ for Taurus.
Even though the Schmidt law is a useful tool to understand the star formation inside the cloud, it does not completely describe  the whole process. For this purpose, we considered the function that represents the cumulative number of sources above a certain extinction $A_{K}$ as a function of that extinction, shown in Fig.~\ref{fig:N(Ak)}. Even though the Schmidt law predicts that the surface density of protostars steeply grows with extinction, the effective number of protostars decreases as the extinction increases, similarly to what is observed for the Orion molecular cloud in \citet{Lada2013}. 

The total number of protostars is defined as \citep{Lada2013}: 
\begin{equation}\label{Eq:N(Ak)}
N_{\mathrm{YSO}} = \int \Sigma_{\mathrm{YSO}}(A_{K}) dS = \int \Sigma_{\mathrm{YSO}}(A_{K}) \lvert S'(>A_{K}) \rvert dA_{K}.
\end{equation}
Indeed, the number of protostars at a given extinction level is the product of the area $S(A_K)$ encompassing that extinction level, and $\Sigma_{\mathrm{YSO}}(A_K)$. The total number of protostars is given by the integral of this product over all extinction in the cloud. Changing the integration variable from $S$ to $A_K$ gives Eq.~\eqref{Eq:N(Ak)}.  
We already studied the function $S(>A_{K})$ (Fig.~\ref{fig:S(Ak)}) and we concluded that its trend is $\propto A_{K}^{-2}$ and therefore $S'(>A_{K}) \propto A_{K}^{-3}$. 
Thus, the number of protostars formed within the cloud does not depend only on the surface density of the gas but also on the area function and its derivative of the integral area function, i.e. the (unnormalized) PDF. 
As \citet{Lombardi2015} showed, the PDFs of a sample of nearby molecular clouds, including Perseus, follow roughly the same profile, with slopes in logarithmic binning between $\sim-2$ and $\sim-4$, which correspond to slopes between $\sim -3$ and $\sim -5$ in the linear binning that was done here.
Therefore, it is conceivable to assume that, if the clouds have the same PDF and the same Schmidt law, then the function that describes the number of protostars above a certain extinction (normalized to the total number of protostars) is roughly the same for all the clouds.

\section{Conclusions \label{Conclusions}}
Our main results are the following ones:
\begin{itemize}
\item We produced optical depth and temperature maps of the Perseus molecular cloud, obtained using the data from the \emph{Herschel} and \emph{Planck} satellites. The  maps have a $36 \,\mathrm{arsec}$  resolution for \emph{Herschel} observations and a $5\, \mathrm{arcmin}$ resolution elsewhere. 
\item We calibrated the optical depth maps using 2MASS/\textsc{NICEST}\ extinction data and we obtained extinction maps at the resolution of \emph{Herschel} with a dynamic range $1\times10^{-2} \mathrm{mag}$ to $20 \,\mathrm{mag}$ of $A_{K}$. In particular, we evaluated the ratio $C_{2.2}/\kappa_{850} = 3620 \pm 252$, i.e. the ratio between the \SI{2.2}{\um} extinction coefficient and of the \SI{850}{\um} opacity.
\item We studied the cumulative and differential area functions of the data, and we showed that starting from $A_{K} \simeq 0.1 \,\mathrm{mag}$ the cumulative area function follows a power law with index $\simeq -2$.
\item We used WISE data to improve current YSO catalogs based mostly on \emph{Spitzer} data and we built an up-to-date selection of Class~I/0 protostars.
\item We evaluated the local Schmidt law, $\Sigma_{\mathrm{YSO}} \propto A_{K}^{\beta}$, using the \emph{Herschel/Planck} maps and the new object selection. We found that $\beta = 2.4 \pm 0.6$.
\item We showed that the Schmidt law does not completely describe the whole star formation process. Indeed, the total number of protostars effectively formed within a cloud depends on the surface area function $S(>A_K)$ and its derivative. 
\end{itemize}

\begin{acknowledgements}
We thank the anonymous referee for comments that improved the manuscript. We also wish to thank Josefa Grossscheld and Paula Stella Teixeira for the helpful discussions.    
This publication would not have been possible without the data products from the {\it Herschel} satellite and the Wide-field Infrared Survey Explorer. {\it Herschel} is an ESA space observatory with science instruments provided by European-led Principal Investigator consortia and with important participation from NASA. WISE is a joint project of the University of California, Los Angeles, and the Jet Propulsion Laboratory/California Institute of Technology, funded by the National Aeronautics and Space Administration. This paper is also based on observations obtained with \emph{Planck} (http://www.esa.int/Planck), an ESA science mission with instruments and contributions directly funded by ESA Member States, NASA, and Canada. Besides, it makes use of data products from the Two Micron All Sky Survey, which is a joint project of the University of Massachusetts and the Infrared Processing and Analysis Center/California Institute of Technology, funded by the National Aeronautics and Space Administration and the National Science Foundation. This research has made use of: the SIMBAD database, operated at CDS,
Strasbourg, France; the VizieR catalogue
access tool, CDS, Strasbourg, France; Astropy, a community-developed core Python package for Astronomy
\citep{2013A&A...558A..33A}; TOPCAT, an
interactive graphical viewer and editor for tabular data
\citep{2005ASPC..347...29T}. \\

\end{acknowledgements}

\appendix
\section{Hidden layers of multiple layer figures}
In this section we provide a "flat" version of the hidden layers of multiple layer figures.

\begin{figure}
\includegraphics[width=\hsize]{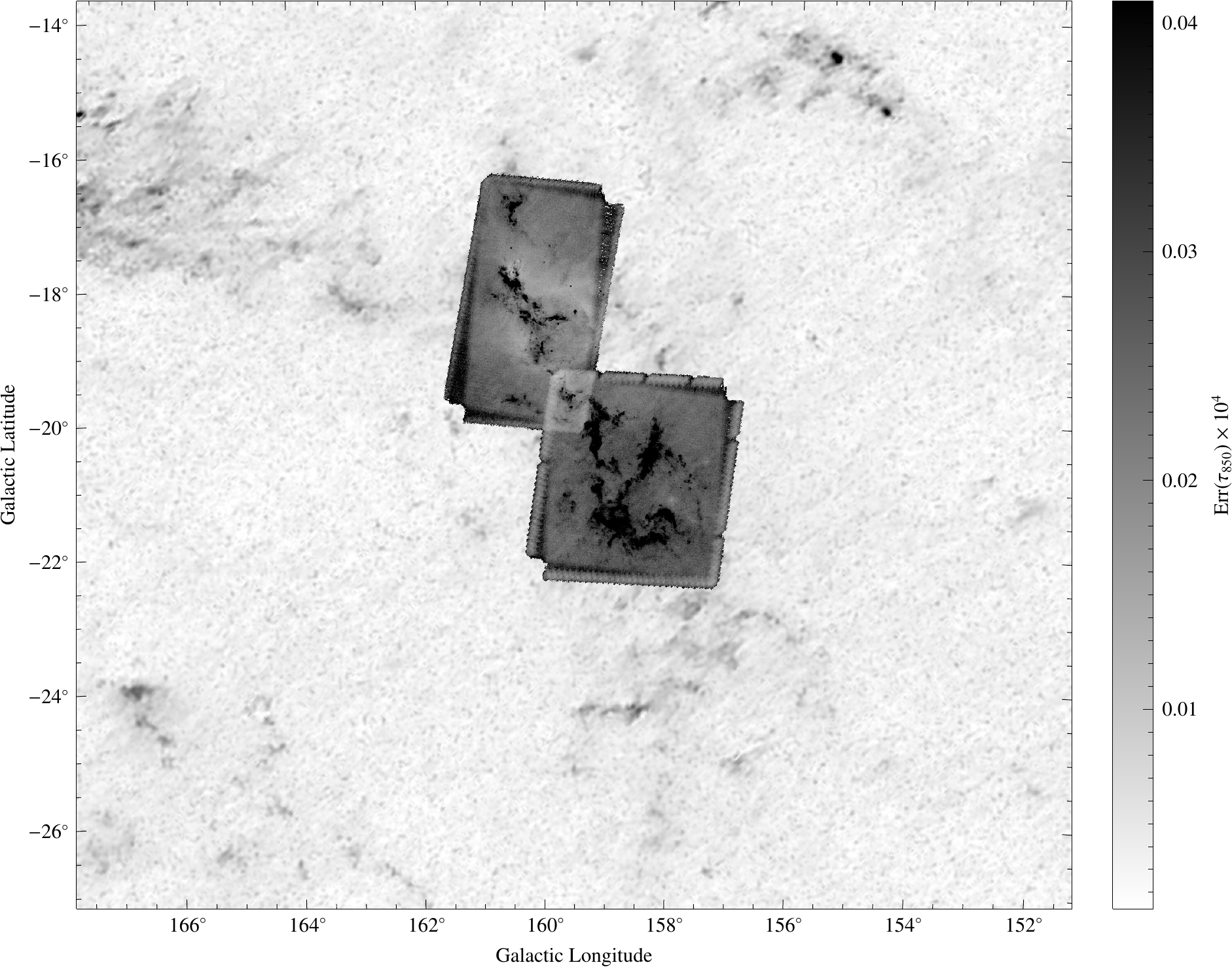}
\caption{Error on the optical depth map (see fig. \ref{fig:opac}). \label{fig:03b}}
\end{figure}

\begin{figure}
\includegraphics[width=\hsize]{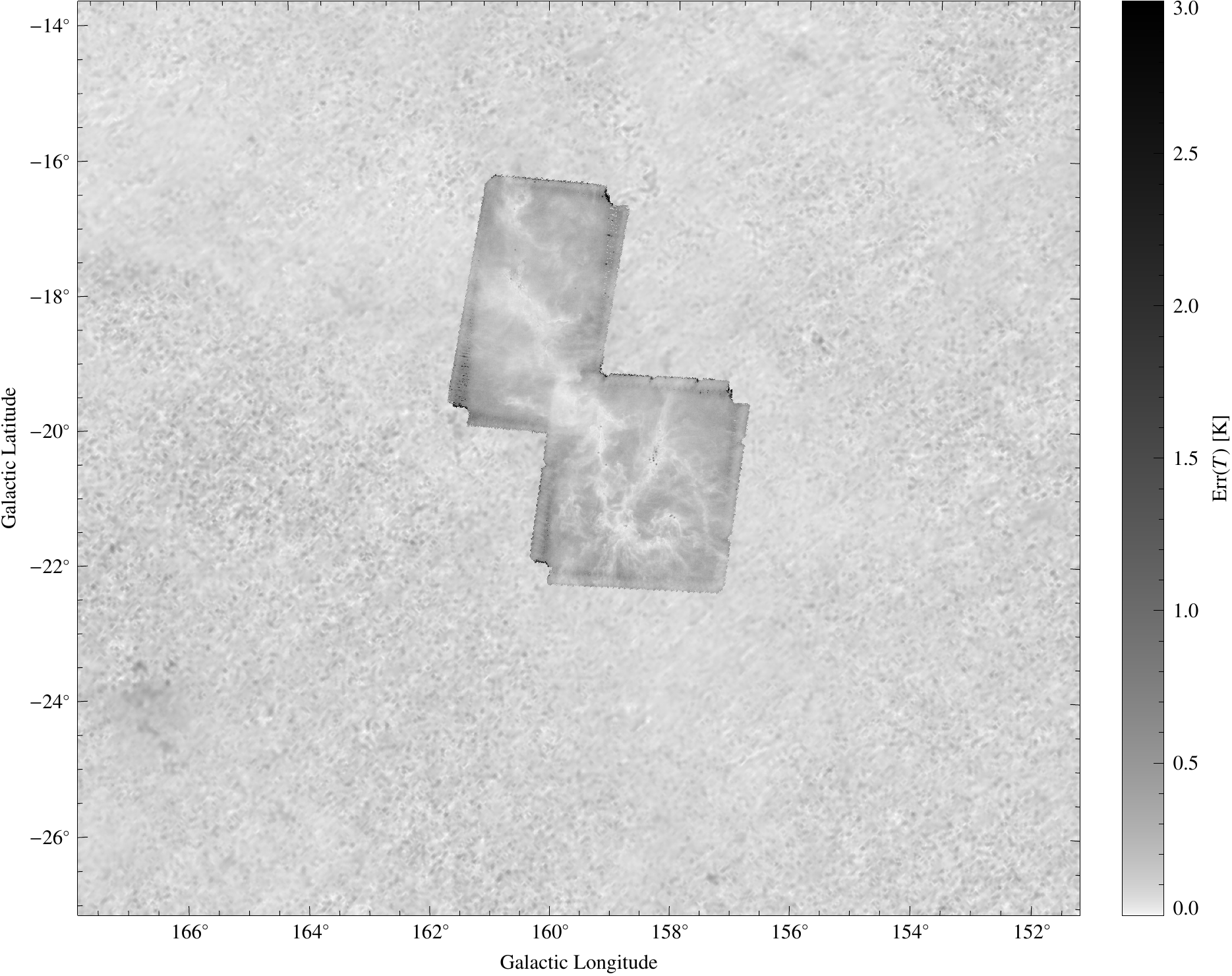}
\caption{Error on the effective dust temperature map (see fig. \ref{fig:temp}). \label{fig:03d}}
\end{figure}

\begin{figure}
\includegraphics[width=\hsize]{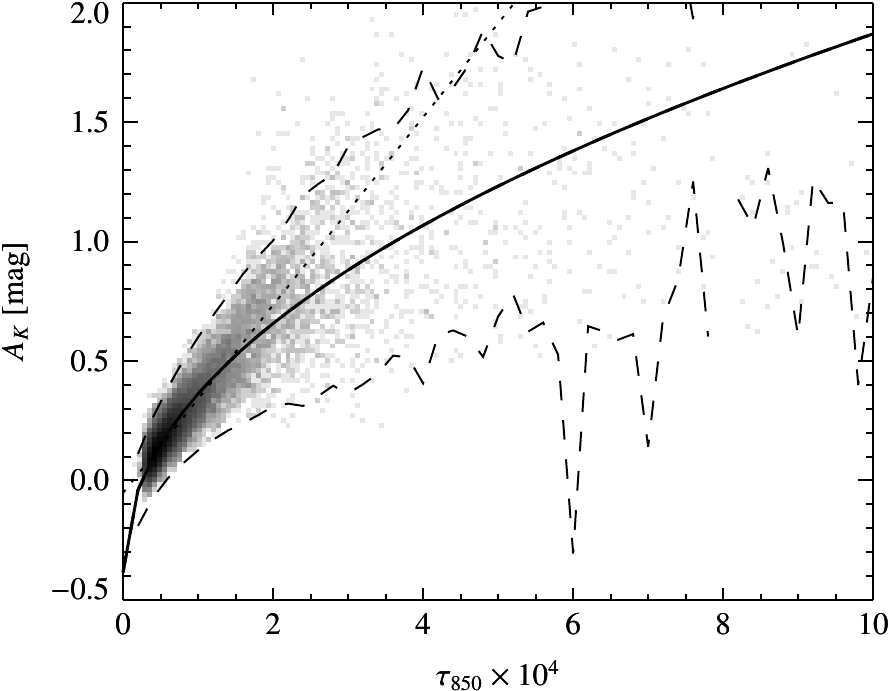}
\caption{Relation between 353 GHz optical depth and the infrared \textsc{NICEST} extinction map, for $\tau_{850}\times10^4 \le 10$. The relation deviates from the linearity for high extinction values, as shown by the fit (solid line).The dashed lines represent the expected $3\sigma$ region. We also report in this plot the linear fit shown in Fig.~\ref{fig:fit_lin} (dotted line).\label{fig:05b-nicest}}
\end{figure}

\section{Potential new star cluster}
It is worth mentioning that while analyzing the spatial distribution of the WISE excess sources, we noticed a small group slightly off-set from the cloud, whose center is located approximately at $l=162.2^{\circ}, \, b = -19.4^{\circ}$. We found no reference in the literature regarding this potential new group of young stars. The group contains fourteen members spread across an area of about 1 square degree. None of the excess sources is  present in the Hipparcos catalog, therefore it is not possible to give an estimate of its distance or its potential relation with the Perseus molecular cloud. Follow up observations for this new group of excess sources is needed. The Gaia mission is expected to dramatically improve the distance determination for faint stars, which will allow for a better understanding of the origins and relevance of this group to the history of the star formation in the Perseus region.

\section{Catalog of the sources used to evaluate the Schmidt law}
In this section we provide the list of protostars used to evaluate the Schmidt law. We report the position for all the sources and the WISE magnitudes for the sources detected by both \emph{Spitzer} and WISE.

\onecolumn
\tablecaption{Catalog of the protostars used to determine the Schmidt law. Y15 stands for \citet{Young2015}, G09 for \citet{Gutermuth2009}, GMM09 for \cite{Gutermuth2008}, E09 for \cite{Evans2009} and S14 for \citet{Sadavoy2014}.}
\begin{supertabular}{lllllll}
RA J2000 [deg] & DEC J2000 [deg] & W1 [mag] & W2 [mag]  & W3 [mag]  & W4 [mag] & Ref. \\
\hline
			& 			& 			&		  						&						&		& \\
55.8726234 & 31.8720837 & 12.18  $\pm$	0.03 &	10.55	$\pm$ 0.02  &	6.66 $\pm$  0.02 &	7.5 $\pm$	0.2	& Y15 \\
55.9000816 & 31.8358345 &                    &                      &                    &                  & Y15 \\
55.9382095 & 32.0662766 &             &                            &                    &                   & Y15 \\
55.9625816 & 32.0522499 &            &                            &                    &                    & Y15 \\
55.9868317 & 32.0513039 &            &                            &                    &                    & Y15 \\
56.0109596 & 32.0331955 &            &                            &                    &                    & Y15 \\
56.0383339 & 32.0438347 & 10.94  $\pm$	0.04 &	9.66  $\pm$  0.07 & 6.6 $\pm$ 0.1 & 7.1 $\pm$ 0.1&  Y15   \\                                                 
56.0540848 & 32.0265274 &                                 &                    &       &                  & Y15 \\
56.0889587 & 31.9923897 &                                 &                   &       &                   & Y15 \\
56.1035004 & 32.2301102 &                                 &                    &       &                  & Y15 \\
56.1262512 & 32.1931114 &                                 &                    &       &                  & Y15 \\
56.1805000 & 32.0254173 & 8.63  $\pm$ 0.03  &	6.93   $\pm$ 0.03 &	4.4   $\pm$ 0.07 &	8.5  $\pm$	0.3		& Y15 \\
56.1831665 & 32.0267220 &                                 &                    &                  &         & Y15 \\
56.3075829 & 32.2027779 &                                 &                    &                  &         & Y15 \\
52.1356659 & 31.1847496 & 13.47   $\pm$ 0.04  &	11.88	 $\pm$ 0.04 &	8.92  $\pm$	0.07 &	8.0  $\pm$	NaN     & Y15 \\
52.1437073 & 31.0141945 &                                 &                    &       &                    & Y15 \\
52.1545410 & 31.2252235 &                                 &                   &           &                 & Y15    \\
52.1615829 & 31.3018322 &                       &                  &                  &       & Y15    \\
52.1629181 & 31.1004734 &                       &                  &                  &       & Y15    \\
52.1654587 & 31.2921944 & 11.43  $\pm$ 0.05 &	9.66  $\pm$	0.04 &	6.96  $\pm$	0.09 &	8.0  $\pm$	0.3 &  Y15 \\
52.1692924 & 31.2990265 &                    &                  &                  &                  & Y15     \\
52.1803322 & 31.2924709 &                    &                  &                  &                  & Y15 \\
52.2032089 & 31.2691116 & 13.50  $\pm$	0.03 &	11.99	 $\pm$ 0.03	& 9.2  $\pm$	0.1 &	8.0  $\pm$	NaN &  Y15  \\
52.2135849 & 31.2942486 &                    &                    &                    &                & Y15   \\
52.2314568 & 31.2435284 &                    &                    &                  &              & Y15  \\
52.2358322 & 31.1269455 &                    &                   &               &                  &   Y15   \\
52.2383766 & 31.2386398 &                       &                  &                  &             & Y15    \\
52.2434578 & 31.3715286 &                         &                &                  &           & Y15   \\
52.2467918 & 31.3423615 &                        &                 &                  &             & Y15    \\
52.2471657 & 31.2635269 &                       &                 &                  &              & Y15   \\
52.2522926 & 31.2002239 &                      &                 &                  &               & Y15   \\
52.2565002 & 31.3390560 &                      &                  &                  &              & Y15   \\
52.2638741 & 31.3873882 &                      &                  &                  &               & Y15   \\
52.2657509 & 31.2677212 & 6.71  $\pm$	0.06 &	4.93  $\pm$	0.06 &	1.26  $\pm$	0.01 &	-1.71 $\pm$	0.002	&	Y15 \\
52.2669182 & 31.2462502 &                      &                  &                 &             & Y15   \\
52.2706261 & 31.3439999 &                      &                  &                  &          &  Y15    \\
52.2824173 & 31.3659172 &                      &                  &                  &             & Y15   \\
52.2873764 & 31.3822498 &                      &                  &                  &            & Y15   \\
52.2879181 & 31.3579731 &                      &                  &                  &             & Y15   \\
52.2937088 & 31.2252789 &                      &                  &                  &            & Y15   \\
52.2943764 & 31.2277775 &                      &                  &                  &            & Y15    \\
52.2944984 & 31.3057213 &                      &                  &                  &            & Y15   \\
52.2969170 & 31.3087215 &                      &                  &                  &            &  Y15   \\
52.2995415 & 31.3575001 &                      &                  &                  &            & Y15   \\
52.3002510 & 31.2181664 &                      &                  &                  &            & Y15   \\
52.3002510 & 31.2171402 &                      &                  &                  &            &  Y15  \\
52.3040428 & 31.3039722 & 10.38	 $\pm$	0.04 &	7.61  $\pm$	0.09 &	5.65  $\pm$	0.07 &	7.5  $\pm$	0.15 & Y15 \\
52.3064156 & 31.2328339 &                      &                 &                  &            &     Y15   \\
52.3215408 & 31.4629173 &                     &                  &                  &            &   Y15    \\
52.3260841 & 31.3888607 &                     &                  &                  &            &   Y15    \\
52.3277931 & 31.3382511 &                     &                  &                  &            &   Y15    \\
52.3335838 & 31.4020824 &                     &                  &                  &            &   Y15     \\
52.3351669 & 31.3094997 &                     &                  &                  &            &   Y15     \\
52.3503761 & 31.3326683 &                     &                  &                  &            &   Y15     \\
51.3313332 & 30.5733891 &                     &                  &                  &            &   Y15     \\
51.3429985 & 30.7538605 &                     &                  &                  &            &   Y15     \\
51.4009171 & 30.7543602 &                     &                  &                  &            &   Y15     \\
51.4020424 & 30.7561665 &                     &                  &                  &            &   Y15     \\
51.4117928 & 30.7350559 &                     &                  &                  &            &   Y15     \\
51.4129982 & 30.7328339 &                     &                  &                  &            &   Y15     \\
51.6561241 & 30.2578049 &                     &                  &                  &            &   Y15     \\
51.9128342 & 30.2175274 &                     &                  &                  &            &   Y15     \\
51.9301262 & 30.2080288 &                     &                  &                  &            &   Y15     \\
51.9486237 & 30.2012501 & 7.4	 $\pm$ 0.2 &	6.07  $\pm$	0.03	 &3.46	 $\pm$	0.01	& 1.64  $\pm$	0.02 & Y15 \\
52.0016251 & 30.1336937 &                     &                  &                  &            &   Y15     \\
52.1437073 & 31.0141945 &                     &                  &                  &            &   Y15     \\
52.1887512 & 31.0949726 & 13.81  $\pm$ 0.02 &	11.63  $\pm$	0.02 &	8.89  $\pm$	0.08 &	8.7  $\pm$	NaN  & Y15 \\
52.2752075 & 30.5108891 &                        &                  &                  &            &   Y15     \\
52.3478317 & 31.5581932 &                        &                  &                  &            &   Y15     \\
52.4659157 & 31.6516666 &                        &                  &                  &            &   Y15     \\
52.5630836 & 30.3970547 &                        &                  &                  &            &   Y15     \\
52.5935402 & 31.5445824 &                        &                  &                  &            &   Y15     \\
52.6130829 & 30.4749451 & 10.8  $\pm$	0.03	 &9.94  $\pm$	0.02 &	7.39  $\pm$	0.04  &	7.4  $\pm$	0.1 &   Y15  \\
52.6361656 & 30.4407215 &                        &                  &                  &            &   Y15   \\
52.8112488 & 30.8320560 & 9.93	 $\pm$  0.03 &	8.82  $\pm$	0.03 &	6.39  $\pm$	0.07 &	6.0  $\pm$		0.1  & Y15\\
52.8374176 & 30.7583618 &       &                   &                  &                      &   Y15    \\
53.0748329 & 30.8298626 &       &                   &                  &                       &   Y15    \\
53.1215401 & 31.0446663 &       &                   &                  &                       &   Y15    \\
53.2410011 & 31.1023064 &       &                   &                  &                       &    Y15   \\
53.2767067 & 31.1346111 &       &                   &                  &                       &   Y15   \\
53.2898331 & 31.0920010 &       &                   &                &                       &   Y15    \\
53.3035011 & 31.3567219 & 7.28  $\pm$	0.04 &	5.73  $\pm$	0.04 &	3.11  $\pm$	0.01 &	0.93  $\pm$		0.02 & Y15 \\
53.3074989 & 31.3348064 &       &                   &                  &                      &   Y15  \\
53.3099174 & 31.1196938 &       &                   &                  &                       &   Y15    \\
53.3185005 & 31.1145840 &       &                   &                  &                       &   Y15    \\
53.3193741 & 31.1320000 &       &                   &                  &                       &   Y15    \\
53.3243752 & 31.1588612 &       &                   &                  &                       &   Y15   \\
53.3346672 & 31.1226387 & 13.55  $\pm$	0.02 &	10.19  $\pm$ 0.02 &	8.2  $\pm$ 0.1 &	8.7  $\pm$	NaN   & Y15\\
53.3637085 & 31.1194992 &       &                   &                  &                      &   Y15     \\
55.4212074 & 31.8012772 &       &                   &                  &                      &   Y15     \\
55.4821243 & 31.8031654 &       &                   &                  &                      &   Y15     \\
55.4944572 & 31.8059444 &       &                   &                  &                      &   Y15     \\
55.5090408 & 31.8005829 &       &                   &                  &                      &   Y15     \\
55.7335434 & 31.9457760 &       &                   &                  &                      &   Y15     \\
56.1472511 & 32.4770012 &       &                   &                  &                      &   Y15     \\
56.7726250 & 32.7190285 & 9.6	  $\pm$	0.02 &	8.22  $\pm$	 0.02 &	5.72  $\pm$	0.04 &	8.9  $\pm$	0.5 & Y15 \\
56.9232483 & 32.8622475 &       &                   &                  &                       &   Y15   \\
52.1438332 & 31.1181946 &       &                   &                  &                       &   G09     \\
52.3001671 & 31.2172222 &       &                   &                  &                       &   G09     \\
52.3428345 & 31.2318058 &       &                   &                  &                       &   G09     \\
52.2390404 & 31.237833  &       &                   &                  &                       &   G09     \\
52.203167  & 31.2691383 & 13.50	 $\pm$ 0.03	 & 11.99  $\pm$	0.03 &	9.25  $\pm$	0.13 &	8.6  $\pm$	NaN      & G09 \\
52.1802902 & 31.2925282 &       &                  &               &                       &   G09     \\
52.2034988 & 31.2982502 &     &                  &                 &                       &   G09     \\
52.237957  & 31.319973  &     &                   &                  &                       &   G09     \\
52.3048325 & 31.3303623 &     &                   &                  &                       &   G09     \\
52.2471237 & 31.3356113 &     &                   &                  &                       &   G09    \\
52.2705841 & 31.3440571 &     &                   &                  &                       &   G09    \\
52.2378769 & 31.3569717 &     &                   &                  &                       &   G09    \\
52.2872925 & 31.3823051 &     &                   &                  &                       &   G09    \\
52.3814163 & 31.4243889 &     &                   &                  &                       &   G09    \\
55.9001236 & 31.8358345 &     &                   &                  &                       &   G09   \\
56.0540848 & 32.0265274 &     &                   &                  &                       &   G09    \\
56.0098763 & 32.027916  &     &                   &                  &                       &   G09    \\
55.9965401 & 32.0474701 &     &                   &                  &                       &   G09   \\
55.9382095 & 32.0662766 &     &                   &                  &                       &   G09    \\
55.9778328 & 32.075695  &     &                   &                  &                       &   G09    \\
56.1262512 & 32.1931381 &     &                   &                  &                       &   G09    \\
56.2982101 & 32.4647789 &     &                   &                  &                       &   G09    \\
52.2416000 & 31.3478000 &       &                 &                  &                       &   GMM09  \\
52.2879000 & 31.3849000 &    	&   		  	 &  		       &	  	      	          &    GMM09   \\
55.9791985 & 32.0175018 &       &                   &                 &                       &   E09   \\
55.9886017 & 32.0131989 &       &                   &                &                       &   E09   \\
55.9855003 & 32.0147018 &       &                   &                  &                       &   E09   \\
55.9902    & 32.0124016 &       &                   &                  &                       &   E09   \\
55.9908981 & 32.0533981 &       &                   &                  &                       &   E09   \\
55.9975014 & 32.0098991 &       &                   &                  &                       &   E09   \\
55.9985008 & 32.0317001 & 7.5	 $\pm$ 0.1  &	6.21  $\pm$	0.03 &	4.10  $\pm$  0.03 & 6.12  $\pm$	0.05 &  E09 \\
56.0890007 & 31.9923992 &                   &       &                  &                        & E09   \\
55.5444984 & 31.7849007 & 13.09  $\pm$	0.03 &	11.93  $\pm$	0.05 &	10.48  $\pm$	0.04 &	8.0  $\pm$	NaN	 &	E09 \\
53.196701  & 30.9878998 & 8.79	 $\pm$	0.02 &	7.64  $\pm$	0.02  &	5.78  $\pm$	0.08 &	8.0  $\pm$	NaN  & E09  \\
51.9094429 & 30.2329445 & 9.58   $\pm$ 0.02 & 8.27   $\pm$ 0.02 & 6.51  $\pm$ 0.03  & 8.7  $\pm$ 0.5 & E09  \\
51.3975    & 30.7589    &                  &                  &                  &            &   S14   \\
52.2575    & 31.2594    &                   &                  &                 &            &   S14     \\
52.3283    & 31.3867    &                   &                  &                  &            &   S14     \\
53.3387    & 31.1242    &                   &                  &                  &            &   S14     \\
51.2893562 & 30.7727299 & 8.43  $\pm$  0.02  & 6.92   $\pm$ 0.02 & 4.2  $\pm$ 0.1 & 8.6  $\pm$ 0.5  & New   \\
55.5620079 & 31.7993259 & 13.94  $\pm$ 0.02  & 12.33  $\pm$ 0.03 & 9.92  $\pm$ 0.04 & 7.3  $\pm$ 0.2 & New  \\
\end{supertabular}
\twocolumn

\bibliography{bibliografia}
\bibliographystyle{aa}

\end{document}